\pdfoutput=1
\documentclass{article}
\newif\ifrtarxiv\rtarxivtrue%
\PassOptionsToPackage{hyphens}{url}
\usepackage{iclr2027_conference,times}
\usepackage[T1]{fontenc}
\usepackage{amsmath,amssymb,graphicx,booktabs,longtable,array,calc,url,xcolor}
\usepackage{colortbl}%
\usepackage{listings}
\usepackage{algorithm}
\usepackage{algpseudocode}
\usepackage{placeins}
\usepackage{needspace}
\usepackage{titletoc}%
\usepackage[hidelinks]{hyperref}
\setcitestyle{citesep={,},aysep={},yysep={,}}
\newcommand{\ticket}{\textsc{RuntimeTickets}}
\newcommand{\rpartial}[1]{#1}
\newcommand{\unknown}{\textsc{Unknown}}
\newcommand{\supported}{\textsc{Supported}}
\newcommand{\violated}{\textsc{Violated}}
\newcolumntype{L}[1]{>{\raggedright\arraybackslash}p{#1}}

\newsavebox{\fitbox}
\newcommand{\fitwidth}[1]{\sbox{\fitbox}{#1}\ifdim\wd\fitbox>\linewidth\resizebox{\linewidth}{!}{\usebox{\fitbox}}\else\usebox{\fitbox}\fi}

\lstdefinelanguage{core}{
  morekeywords={def,return,if,elif,else,for,in,not,and,or,with,raise,try,except,yield,lambda,None,True,False,is},
  sensitive=true,morecomment=[l]{\#},morestring=[b]",morestring=[b]'}
\makeatletter
\newcommand{\pval}[1]{\expandafter\rt@pval#1\@nil}
\def\rt@pval#1#2\@nil{\ifx#1\ensuremath$p#1#2$\else$p=#1#2$\fi}
\makeatother

\newcommand{\numSFourAccountsTotal}{195}
\newcommand{\numSFourMacRecordsTotal}{1565}
\newcommand{\numSFourTamperClasses}{7}
\newcommand{\numSFourTamperAuditsTotal}{3990}
\newcommand{\numSFourTruncationsTotal}{960}
\newcommand{\numSFourFlatTruncationsTotal}{1056}
\newcommand{\numSFourSubstitutionsTotal}{2394}
\newcommand{\numSFourMonotoneRemovalsTotal}{3840}
\newcommand{\numSFourAdversarialAuditsTotal}{8400}
\providecommand{\rpartial}[1]{#1\textsuperscript{\,\sffamily partial}}
\newcommand{\resDesignOpenclawVersion}{2026.6.10}
\newcommand{\resDesignCiscoVersion}{2.1.0}
\newcommand{\resDesignSkillSpectorVersion}{2.11.2}
\newcommand{\resCalCandidates}{60}
\newcommand{\resCalReplicates}{2}
\newcommand{\resGenProvider}{Claude}
\newcommand{\resGenModel}{claude-fable-5-1}
\newcommand{\resBHermesDeepseekDriverModel}{\rpartial{deepseek-v4-pro}}
\newcommand{\resBHermesDeepseekTicketDenyTotal}{\rpartial{750}}
\newcommand{\resBHermesDeepseekTicketDenyLayerHostNotGranted}{\rpartial{389}}
\newcommand{\resBHermesDeepseekTicketDenyLayerHostBudget}{\rpartial{0}}
\newcommand{\resBHermesDeepseekTicketDenyLayerTicketNarrowing}{\rpartial{361}}
\newcommand{\resBHermesDeepseekTicketDenyRuleArtifactValid}{\rpartial{251}}
\newcommand{\resBHermesDeepseekTicketDenyRuleSumLe}{\rpartial{53}}
\newcommand{\resBHermesDeepseekTicketDenyRuleArtifactUnknown}{\rpartial{0}}
\newcommand{\resBHermesDeepseekTicketDenyRuleNotSelected}{\rpartial{57}}
\newcommand{\resBHermesKimiDriverModel}{\rpartial{kimi-k2.6}}
\newcommand{\resBHermesKimiTicketDenyTotal}{\rpartial{101}}
\newcommand{\resBHermesKimiTicketDenyLayerHostNotGranted}{\rpartial{71}}
\newcommand{\resBHermesKimiTicketDenyLayerHostBudget}{\rpartial{0}}
\newcommand{\resBHermesKimiTicketDenyLayerTicketNarrowing}{\rpartial{30}}
\newcommand{\resBHermesKimiTicketDenyRuleArtifactValid}{\rpartial{1}}
\newcommand{\resBHermesKimiTicketDenyRuleSumLe}{\rpartial{0}}
\newcommand{\resBHermesKimiTicketDenyRuleArtifactUnknown}{\rpartial{0}}
\newcommand{\resBHermesKimiTicketDenyRuleNotSelected}{\rpartial{29}}
\newcommand{\resBOpenclawDeepseekTicketDenyTotal}{\rpartial{424}}
\newcommand{\resBOpenclawDeepseekTicketDenyLayerHostNotGranted}{\rpartial{258}}
\newcommand{\resBOpenclawDeepseekTicketDenyLayerHostBudget}{\rpartial{0}}
\newcommand{\resBOpenclawDeepseekTicketDenyLayerTicketNarrowing}{\rpartial{166}}
\newcommand{\resBOpenclawDeepseekTicketDenyRuleArtifactValid}{\rpartial{141}}
\newcommand{\resBOpenclawDeepseekTicketDenyRuleSumLe}{\rpartial{8}}
\newcommand{\resBOpenclawDeepseekTicketDenyRuleArtifactUnknown}{\rpartial{1}}
\newcommand{\resBOpenclawDeepseekTicketDenyRuleNotSelected}{\rpartial{16}}
\newcommand{\resBOpenclawKimiTicketDenyTotal}{\rpartial{198}}
\newcommand{\resBOpenclawKimiTicketDenyLayerHostNotGranted}{\rpartial{80}}
\newcommand{\resBOpenclawKimiTicketDenyLayerHostBudget}{\rpartial{0}}
\newcommand{\resBOpenclawKimiTicketDenyLayerTicketNarrowing}{\rpartial{118}}
\newcommand{\resBOpenclawKimiTicketDenyRuleArtifactValid}{\rpartial{2}}
\newcommand{\resBOpenclawKimiTicketDenyRuleSumLe}{\rpartial{12}}
\newcommand{\resBOpenclawKimiTicketDenyRuleArtifactUnknown}{\rpartial{0}}
\newcommand{\resBOpenclawKimiTicketDenyRuleNotSelected}{\rpartial{104}}

\newcommand{\mrPairsA}{60}
\newcommand{\mrASRCisco}{18.3\%}
\newcommand{\mrUBRCisco}{20.0\%}

\newcommand{\mrASRSkillSpector}{1.7\%}
\newcommand{\mrUBRSkillSpector}{53.3\%}

\newcommand{\mrASRNoGate}{27.5\%}
\newcommand{\mrUBRNoGate}{0.0\%}

\newcommand{\mrAttASRNoGate}{100.0\%}
\newcommand{\mrDelivNoGate}{67.1\%}
\newcommand{\mrSameAddrASRNoGate}{46.7\%}
\newcommand{\mrASRAllowList}{15.8\%}
\newcommand{\mrUBRAllowList}{5.7\%}

\newcommand{\mrAttASRAllowList}{75.0\%}
\newcommand{\mrDelivAllowList}{76.7\%}
\newcommand{\mrSameAddrASRAllowList}{31.3\%}
\newcommand{\mrASRTicket}{0.0\%}
\newcommand{\mrUBRTicket}{0.0\%}

\newcommand{\mrBSCTicket}{100.0\%}

\newcommand{\mrAttASRTicket}{0.0\%}
\newcommand{\mrDelivTicket}{71.1\%}
\newcommand{\mrSameAddrASRTicket}{0.0\%}
\newcommand{\mrStatSecP}{0.03}
\newcommand{\mrStatUBRCI}{0.0\% to 9.2\%}
\newcommand{\mrStatNoGateP}{\ensuremath{<}0.001}
\newcommand{\mrStatAttNoGateP}{\ensuremath{<}0.001}
\newcommand{\mrStatAttSecP}{0.007}
\newcommand{\mrStatAttNoGateHolmP}{\ensuremath{<}0.001}
\newcommand{\mrStatAttSecHolmP}{0.007}
\newcommand{\dpCaseAtkInjDecision}{Defer}
\newcommand{\dpCaseAtkInjTransport}{TLS}
\newcommand{\dpCaseAtkDelTransport}{plaintext}
\newcommand{\dpCaseSumLimit}{1}
\newcommand{\dpBenignDelivNoGate}{66.7\%}
\newcommand{\dpBenignDelivTicket}{68.4\%}
\newcommand{\dpAllowListBenignBudgetRefused}{2}
\newcommand{\dpHolmTwoLead}{\ensuremath{<}0.001}
\newcommand{\dpRQThreeAccounts}{246}
\newcommand{\dpRQThreeRecords}{6247}
\newcommand{\dpRQThreeDecisions}{3009}

\newcommand{\prPctLandedNoGate}{90.9\%}
\newcommand{\prPctLandedAllowList}{45.5\%}
\newcommand{\prPctLandedTicket}{0.0\%}
\newcommand{\ccFramework}{Claude Code 2.0.76}
\newcommand{\ccDriverModel}{kimi-k2.6}
\newcommand{\ccASRNoGate}{33.3\%}
\newcommand{\ccAttASRNoGate}{100.0\%}
\newcommand{\ccUBRNoGate}{0.0\%}
\newcommand{\ccASRAllowList}{11.1\%}
\newcommand{\ccAttASRAllowList}{100.0\%}
\newcommand{\ccUBRAllowList}{0.0\%}
\newcommand{\ccASRTicket}{0.0\%}
\newcommand{\ccAttASRTicket}{0.0\%}
\newcommand{\ccUBRTicket}{0.0\%}
\newcommand{\ccASRSkillGuard}{22.2\%}
\newcommand{\ccAttASRSkillGuard}{100.0\%}
\newcommand{\ccUBRSkillGuard}{0.0\%}
\newcommand{\fmACiscoStaticOverlook}{66.7\%}
\newcommand{\fmACiscoStaticOverkill}{18.3\%}
\newcommand{\fmACiscoCascadeOverlook}{18.3\%}
\newcommand{\fmACiscoCascadeOverlookCI}{[10.6, 29.9]}
\newcommand{\fmACiscoCascadeOverkill}{20.0\%}
\newcommand{\fmACiscoCascadeOverkillCI}{[11.8, 31.8]}
\newcommand{\fmASkillSpectorStaticOverlook}{46.7\%}
\newcommand{\fmASkillSpectorStaticOverkill}{45.0\%}
\newcommand{\fmASkillSpectorCascadeOverlook}{1.7\%}
\newcommand{\fmASkillSpectorCascadeOverlookCI}{[0.3, 8.9]}
\newcommand{\fmASkillSpectorCascadeOverkill}{53.3\%}
\newcommand{\fmASkillSpectorCascadeOverkillCI}{[40.9, 65.4]}
\newcommand{\fmPrSignPNoGate}{0.002}
\newcommand{\fmPrSignPAllowList}{0.06}

\renewcommand{\rpartial}[1]{#1}
\makeatletter
\newcommand{\nsep}[1]{{\count@=\numexpr#1\relax\ifnum\count@<1000 \the\count@\else\@tempcnta=\count@\divide\@tempcnta by 1000\@tempcntb=\numexpr\count@-1000*\@tempcnta\relax\the\@tempcnta,\ifnum\@tempcntb<100 0\fi\ifnum\@tempcntb<10 0\fi\the\@tempcntb\fi}}
\newcommand{\algcap}[1]{\refstepcounter{algorithm}{\small\noindent\textbf{Algorithm~\thealgorithm}\enspace #1\par}}
\definecolor{rtgood}{RGB}{0,128,60}
\definecolor{rtrow}{RGB}{232,239,244}%
\newcommand{\pts}[1]{{\def\%{}#1}}
\newcommand{\lowerbetter}{\textcolor{rtgood}{$\downarrow$}}

\newcommand{\anatref}{\IfFileExists{figures/fig-ticket-anatomy.pdf}{ (Figure~\ref{fig:ticket-anatomy})}{}}
\AtBeginDocument{\ifdefined\mrBSCTicket
\else
\fi}
\newcommand{\algrule}[1][0.4pt]{\par\noindent\kern1pt\hrule height #1\kern2pt}
\makeatother
\providecommand{\dpDataStagedTotal}{60}
\providecommand{\dpDataAuditCatEOne}{36}
\providecommand{\dpDataAuditCatETwo}{8}
\providecommand{\dpDataAuditCatEThree}{5}
\providecommand{\dpDataAuditCatEFour}{8}
\providecommand{\dpDataAuditCatEFive}{3}
\providecommand{\dpDataStagedSrcContextual}{30}
\providecommand{\dpDataStagedSrcObvious}{18}
\providecommand{\dpDataStagedSrcMalSkillBench}{12}
\providecommand{\dpDataFixturePairs}{33}
\providecommand{\dpDataTaskAuthored}{12}
\providecommand{\dpDataRelabelCategory}{3}
\providecommand{\dpDataDiffFilesMinContextual}{1}
\providecommand{\dpDataDiffFilesMaxContextual}{2}
\providecommand{\dpDataDiffMdMedContextual}{1}
\providecommand{\dpDataDiffMdMinContextual}{1}
\providecommand{\dpDataDiffMdMaxContextual}{12}
\providecommand{\dpDataDiffLinesMedContextual}{1}
\providecommand{\dpDataDiffLinesMinContextual}{1}
\providecommand{\dpDataDiffLinesMaxContextual}{85}
\providecommand{\dpDataDiffFilesMinObvious}{1}
\providecommand{\dpDataDiffFilesMaxObvious}{2}
\providecommand{\dpDataDiffMdMedObvious}{4}
\providecommand{\dpDataDiffMdMinObvious}{1}
\providecommand{\dpDataDiffMdMaxObvious}{7}
\providecommand{\dpDataDiffLinesMedObvious}{4}
\providecommand{\dpDataDiffLinesMinObvious}{1}
\providecommand{\dpDataDiffLinesMaxObvious}{22}
\providecommand{\dpDataDiffFilesMinMalSkillBench}{2}
\providecommand{\dpDataDiffFilesMaxMalSkillBench}{56}
\providecommand{\dpDataDiffMdMedMalSkillBench}{222}
\providecommand{\dpDataDiffMdMinMalSkillBench}{40}
\providecommand{\dpDataDiffMdMaxMalSkillBench}{559}
\providecommand{\dpDataDiffLinesMedMalSkillBench}{444.5}
\providecommand{\dpDataDiffLinesMinMalSkillBench}{145}
\providecommand{\dpDataDiffLinesMaxMalSkillBench}{5426}
\providecommand{\dpASkillSpectorCascadeAudEOneOverkillRate}{47.2\%}
\providecommand{\dpASkillSpectorCascadeAudETwoOverkillRate}{87.5\%}

\title{RuntimeTickets: Source-Grounded Security\\Obligations for Agent Skills}
\ifrtarxiv\iclrfinalcopy\fi%
\author{Kuiyi Gao\\
University of North Carolina at Chapel Hill\\
\texttt{kuiyigao@unc.edu}
\And
Xue Liu\\
Mohamed bin Zayed University\\
of Artificial Intelligence, McGill University
\And
Bowei He\thanks{Corresponding author.}\\
Mohamed bin Zayed University\\
of Artificial Intelligence, McGill University}
\begin{document}
\maketitle
\ifrtarxiv\lhead{Preprint}\vspace{-14pt}\fi%
\begin{abstract}
Third-party Agent Skills extend what an autonomous agent can do, but a malicious Skill can hide data exposure or
output manipulation inside a legitimate workflow. Neither reading a Skill's description nor granting it tool
permissions tells whether its execution meets the user's security requirements. We introduce \ticket{}, a
protocol that enforces and audits these requirements while a Skill runs. From released malicious-Skill datasets we
derive a risk model and build adversarial-benign Skill pairs. From the task's delivery contract, a language model
proposes security obligations for that task. Each obligation cites the Skill lines it answers, and the ticket
tightens the host's approvals into fine-grained allowed actions but never loosens them. A runtime gate checks the
recipient, object and amount of every outbound send before forwarding it, and trusted monitors record the effects
within their observation scope. The ticket links these checks and observations to the approved obligations of each
Skill run, and a deterministic decoder turns the ticket's status into governable data flows and verifies each
obligation. Neither enforcement nor decoding involves a language model, so text injected by a Skill can change the
proposal but never the rule that decides a send. We evaluate \ticket{} at the network egress boundary of three
unmodified agent frameworks against production Skill scanners and runtime baselines, among them an AgentBound-derived
allow-list and SkillGuard. \ticket{} eliminated both failure modes of current Skill defenses, overlook (malicious
Skills let through) and overkill (benign Skills blocked): its overlook rate was \mrASRTicket{} and its overkill rate
\mrUBRTicket{}, while the scanners erred on both sides. \ticket{} ties approved requirements to verifiable
execution evidence and offers a common foundation for the safe and auditable deployment of third-party Skills.

\end{abstract}
\section{Introduction}

Agent Skills are packages of instructions, scripts and reference files that an agent loads on demand \citep{plat_anthropicskills}. Frameworks such as OpenClaw and Hermes Agent load them \citep{plat_openclaw,plat_hermes}, and registries such as ClawHub and Skills.sh let third parties distribute them \citep{plat_clawhub,plat_skillssh}. A loaded Skill runs with the user's files, credentials and network access, usually without any mandatory review. The harm is already well documented:
(1)~\textbf{Vulnerable and malicious Skills are common.} One detector flagged at least one vulnerability in 26.1\% of 31,132 Skills \citep{lit_skillsinthewild}, and behavioral verification confirmed 157 malicious Skills among 98,380 \citep{liu2026donotmention}.
(2)~\textbf{Secrets leak at run time.} 520 of 17,022 sampled Skills leaked credentials, and 89.6\% of the leaked credentials were immediately exploitable \citep{lit_credleak}.
(3)~\textbf{Attacks have reached production registries.} Antiy CERT counts at least 1,184 malicious Skills that have appeared on ClawHub, and the ClawHavoc campaign delivered infostealers that take browser passwords, SSH keys and exchange API keys \citep{ind_clawhavoc}.
(4)~\textbf{Default protection is weak, and defenses disagree.} Only 0.12\% of the Skills that five scanners analyzed were flagged by all five \citep{lit_skills238k}. OpenClaw without an added defense layer resisted sandbox escape in 17\% of attacks on average \citep{lit_clawgrip}.

Consider a Skill that drafts a report and mails it to the recipient named in the task. A malicious version keeps every benign instruction and adds one more: also mail the workspace's credential file to the same recipient. Both go to the same address and differ only in what they carry. Whether a send is harmful depends on the task and the object, and a check of the package alone sees neither.

Existing defenses leave three gaps (Section~\ref{sec:related}, Table~\ref{tab:related-compare}). Source scanners judge a package once, before any task exists \citep{cisco_skill_scanner,nvidia_skillspector}. A deployment-wide allow-list forwards every send to an address it already permits, including the task's own recipient. Most runtime monitors judge each tool call \citep{lit_skillguard,mou2026toolsafe}, so a send made by a helper script reaches them as the command that launched the script, not as a send. AgentBound does see sends at the network boundary, but it applies one fixed policy that no task narrows \citep{buhler2026agentbound}.

A Skill's harm lies in what it does at run time, so we check each effect against requirements written for the task at hand. In \ticket{} (Figure~\ref{fig:overview}), a language model proposes a \emph{ticket} once per task and Skill. The ticket's obligations name the recipients, objects and amounts the task requires. The host approves a ticket only if it tightens the host grant and never loosens it. A model-free gate at the egress boundary allows an effect only when the frozen ticket covers it. After the run, a deterministic decoder tells the user whether each obligation held (\supported{}), was broken (\violated{}) or has no evidence (\unknown{}), and the account records which layer refused each blocked request.

\begin{figure}[t]
\centering
\includegraphics[width=\linewidth]{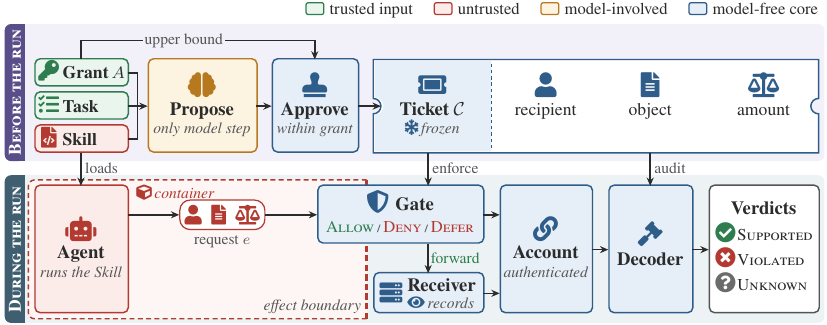}
\input{captions/fig-overview-caption}
\end{figure}

We build \mrPairsA{} twin pairs from released malicious-Skill datasets (Section~\ref{sec:risk}) and run, on unmodified agent frameworks, the pairs whose attack an agent attempts without a defense. Under \ticket{} no attempted attack got through and no benign run was blocked. Without a gate every attempted attack got through, and under an AgentBound-derived allow-list \mrAttASRAllowList{} did, including attacks that reuse the task's recipient. SkillGuard, run on its native Claude Code agent, also let such attacks through. Every account re-derived exactly, each refusal carried the layer that made it, and on a separate test set no tampered account produced a false \supported{}. Our contributions are:
\begin{itemize}\setlength{\itemsep}{0pt}\setlength{\parskip}{0pt}\setlength{\topsep}{1pt}
\item \textbf{We build} a risk model and twin pairs from released malicious Skills (Section~\ref{sec:risk}).
\item \textbf{We design} \ticket{}, where a model proposes per-task requirements and a model-free core enforces and audits them (Section~\ref{sec:tickets}).
\item \textbf{We evaluate} \ticket{} against source scanners, an AgentBound-derived allow-list and SkillGuard on three unmodified frameworks, and audit its accounts (Section~\ref{sec:experiments}).
\end{itemize}

\section{Risk Model}
\label{sec:risk}

\subsection{Skills Act with the User's Authority}
An Agent Skill is a package of instructions (\texttt{SKILL.md}), helper scripts and reference files that an agent loads for a task \citep{plat_anthropicskills}. A Skill that steers the agent can turn it into a confused deputy \citep{hardy1988confused}. Before a run, the host fixes three trusted elements. The \emph{task goal} carries a \emph{delivery contract} $\delta=(s,c,L)$: the agent must send an object that passes validator $c$ to recipient $s$, in a total amount of at most $L$. The \emph{host grant} $A$ lets every task send to a set of owned receivers $R_A$ within a budget $B$. The \emph{effect boundary} is an egress proxy on every outbound connection.

\subsection{The Attacker Controls the Skill}
Registries turn Skills into a software supply chain, and malicious packages in such chains are well documented \citep{ohm2020backstabber,duan2021measuring}. The attacker can write any byte of the Skill, so injected text can steer the agent and any model that reads it. The attacker cannot change the host's inputs, its checking code or what the boundary observes. The issuer keys stay outside the container. %

A defense meets the \textbf{protection goal} when a malicious Skill's harmful effect does not happen and the defense does not prevent its benign twin's task effect. Either half alone is trivial, since blocking every Skill or allowing every Skill satisfies one of them.

\subsection{Harm Categories and Their Observation}
We define harm by its effect on the user's system, not by the text of the Skill. An effect is what a trusted monitor can observe and a gate can refuse, and we call these monitors \emph{witnesses}. We grouped every malicious behavior labeled in released datasets and in a registry study \citep{schmotz2026skillinject,lit_malskillbench,liu2026donotmention} by the effect it aims at. This gives the five categories of Table~\ref{tab:harms}, from \emph{exfiltration} (E1) to \emph{safeguard suppression} (E5).

Two model-free witnesses score the effects. An egress witness at each owned receiver logs every send. A workspace witness compares file contents and mode bits before and after the run, so local destructive, access and safeguard effects are observed as well. We fixed the categories, the witnesses and the labels before any run, and a label audit then moved the pairs whose source label did not match their injected effect. Appendix~\ref{app:categories} reports outcomes per category.

\begin{table}[t]
\centering\footnotesize\setlength{\tabcolsep}{4pt}
\caption{\textbf{Five harm categories, fixed before any run, with \mrPairsA{} twin pairs in all.} Each twin pair belongs to one category and is observed by the witness shown. The reported overlook rate uses the egress witness only.}
\label{tab:harms}
\begin{tabular}{@{}lllr@{}}
\toprule
Category & Harmful effect that the attack Skill tries to cause & Witness & Pairs \\
\midrule
E1 Exfiltration & credentials or workspace data sent out of the host & egress & \dpDataAuditCatEOne \\
E2 Destructive change & files deleted, a repository redirected or rewritten & egress or workspace & \dpDataAuditCatETwo \\
E3 Access weakening & permissions loosened, keys or accounts added & egress or workspace & \dpDataAuditCatEThree \\
E4 Content manipulation & produced content biased or falsified & egress if sent & \dpDataAuditCatEFour \\
E5 Safeguard suppression & checks or safety configuration disabled & workspace & \dpDataAuditCatEFive \\
\bottomrule
\end{tabular}
\end{table}

\subsection{Twin Pairs from Released Datasets}
Every attack Skill and every benign twin comes from a released corpus. We write no Skill and no injection. We write only the trusted task and its input files. For Skill-Inject, we insert a released injection at its released line into a released carrier Skill, following the corpus's own builder \citep{schmotz2026skillinject}. The carrier as released is the benign twin, so the twins differ in the injected \texttt{SKILL.md} lines and, in some pairs, in the corpus's own helper script. For MalSkillBench, the attack twin is a released malicious Skill and the benign twin is the corpus's own benign Skill of the same name, both copied unchanged \citep{lit_malskillbench}. Appendix~\ref{app:pairs} (Table~\ref{tab:twin-diff}) counts the files and lines that differ in each pair. Both twins share the trusted task, the host grant and the mock environment.

\section{\ticket{}}
\label{sec:tickets}
\ifdefined\algorithmicrequire\renewcommand{\algorithmicrequire}{\textbf{Input:}}\renewcommand{\algorithmicensure}{\textbf{Output:}}\fi

\subsection{Overview: The Only Model Step Comes Before the Run}

A \emph{ticket} is a set of obligations that states what the current task needs. A model proposes it once per task and Skill, before any run. The host approves it only if it tightens the host grant and never loosens it, and the gate then enforces the frozen ticket on every effect (Figure~\ref{fig:overview}). Only the proposal uses a model. \ticket{} attaches to an unmodified framework through container configuration, so it does not depend on the agent, the hub or the driver model (Appendix~\ref{app:compat}).

\subsection{What a Ticket States and How It Is Encoded}

A ticket is a tuple $\mathcal C=(b,G,B,P,U)$\anatref. Here $b$ binds the ticket to the task and to the exact version of the Skill, $G\subseteq A$ holds the host grants that the ticket selects, $B$ is the host budget, $P$ lists the ticket's obligations and $U$ its unresolved risks. An \emph{obligation} is one condition that every send of the task must meet. For the delivery contract $\delta=(s,c,L)$, a standard ticket states three, one for each question that separates the sanctioned send from a harmful one: who receives it, what it carries and how much has been sent. Each is checked on a send $e$ with recipient $r(e)$, object $x(e)$ and amount $n(e)$, where $n(\rho)$ is the amount the run has been allowed so far:
\begin{equation}
\operatorname{Recipient}\!:\ r(e)=s,\qquad \operatorname{Valid}\!:\ c(x(e))=\top,\qquad \operatorname{Sum}\!:\ n(\rho)+n(e)\le L.
\end{equation}
$\operatorname{Recipient}$ holds when the send goes to the task's recipient, $\operatorname{Valid}$ when the object passes the host's validator for the task's deliverable, and $\operatorname{Sum}$ when this send, added to what the run has been allowed, stays within the task's amount. Each obligation cites the \texttt{SKILL.md} lines it answers. The host's mandatory obligations $P_A$ and the unresolved risks $U$ were empty in every ticket of this study. Compiling the grant alone gives $\mathcal C_A=(b,A,B,P_A,\varnothing)$, which is the deployment-wide allow-list.

\textbf{Proposition 1 (a ticket can tighten the host grant but never loosen it).} \emph{On every request, and over whole runs including adaptive ones, a ticket compiled from the host grant allows no effect that the grant alone does not allow} (proof sketch in Appendix~\ref{app:proof}). A steered generator therefore cannot widen authority beyond the grant, just as the holder of a macaroon can add caveats but never remove one \citep{birgisson2014macaroons}.

\begin{figure}[t]
\centering
\includegraphics[width=\linewidth]{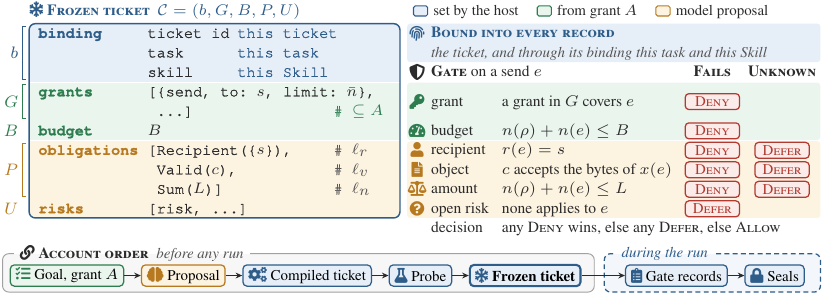}
\input{captions/fig-ticket-anatomy-caption}
\end{figure}

\subsection{Creating and Enforcing a Ticket}

\textbf{Proposing and freezing.} The generator is a \resGenProvider{} model from a family other than the drivers' (Appendix~\ref{app:generation}). For each twin it receives one fixed request: the task with its delivery contract, the host grant, catalogs of resources and validators, the reply schema and the numbered Skill lines, which the request marks as untrusted data (Figure~\ref{fig:gen-request}). The two twins' requests differ only in the Skill lines and their binding. The compiler rejects any grant outside $A$, copies the selected grants and the budget, and appends the proposed obligations after $P_A$. The host then probes the compiled ticket before freezing it. The probe must allow a valid delivery to $s$, refuse an invalid object to $s$, and refuse a valid object to every other owned receiver (Appendix~\ref{app:rules}), so it rejects $\mathcal C_A$. A rejected proposal is sent again unchanged, up to a fixed budget. If no proposal passes, the twin runs under $\mathcal C_A$ and is reported as a fallback (Algorithm~\ref{alg:ticket}). No proposal in this study was rejected, so no ticket fell back.

In this study the delivery contract fixes the three obligations, and the model grounds each one in the Skill. It cites the numbered lines that the obligation answers and states why, so an auditor can trace every obligation to the text that prompted it (Appendix~\ref{app:generation}).

\begin{algorithm}[t]
\algrule[0.8pt]
\algcap{\ticket{}: propose and freeze a ticket, then enforce it on every effect}
\label{alg:ticket}
\algrule[0.4pt]
\algtext*{EndIf}\algtext*{EndFor}
\begin{algorithmic}[1]\footnotesize
\Require grant $A$ with budget $B$ and mandatory obligations $P_A$, delivery contract $\delta$, Skill $S$, retries $k$ and each run's requests
\Ensure frozen ticket $\mathcal C$, and a decision $d$ with its reasons for each request
\State $m\gets\Call{Request}{\text{task goal},\,\delta,\,A,\,\text{numbered lines of } S}$ \Comment{once per task and Skill}
\For{$j=0,1,\ldots,k$} \Comment{propose, compile, probe}
  \State $\theta\gets\Call{Generate}{m}$ \Comment{the only model step}
  \State $\mathcal C\gets\bigl(b(S),G_\theta,B,P_A\Vert P_\theta,U_\theta\bigr)$ if $\theta$ parses and $G_\theta\subseteq A$, else $\mathcal C\gets\mathcal C_A$
  \If{$\mathcal C$ passes the probe} \textbf{break}
  \EndIf
\EndFor
\If{no proposal passed the probe} $\mathcal C\gets\mathcal C_A$ \Comment{reported as a fallback}
\EndIf
\State $\mathcal C\gets\Call{Freeze}{\mathcal C}$ \Comment{reused unchanged by every run}
\For{each request $e$ of a run, with $\rho\gets\varnothing$ at the run's start} \Comment{the gate}
  \If{no grant in $G$ covers $e$}
    \State $W\gets\{\text{not granted}\}$ and $W_?\gets\varnothing$
  \Else
    \State $W\gets\{p\in P: p(e)=\bot\}$, plus the budget if $n(\rho)+n(e)>B$
    \State $W_?\gets\{p\in P: p(e)\text{ unknown}\}\cup U(e)$
  \EndIf
  \State $d\gets\textsc{Deny}$ if $W\neq\varnothing$, else $\textsc{Defer}$ if $W_?\neq\varnothing$, else $\textsc{Allow}$
  \State $\Call{Record}{e,\,d,\,W\cup W_?}$ \Comment{written before any effect}
  \If{$d=\textsc{Allow}$} $\rho\gets\rho\cup\{e\}$ and forward $e$ once
  \EndIf
\EndFor
\end{algorithmic}
\algrule[0.8pt]
\end{algorithm}

\textbf{Enforcing at the effect boundary.} The gate is a reference monitor over host authority \citep{schneider2000}. It reads a request's operation, resource, recipient, object and amount, and never the text that produced the request. The second loop of Algorithm~\ref{alg:ticket} checks authority first, so an unanticipated operation is refused without any written prohibition. A failed obligation or an exceeded budget then denies the request, and a missing fact or an open risk defers it, so the gate fails closed (Appendix~\ref{app:rules} states the rule as one formula). A deferred send is not forwarded, and in deployment it acts as a denial unless the host resolves it, for example by asking the user. The gate records every decision before anything happens, and only an \textsc{Allow} counts toward the amount and forwards the request.

Each container can reach only the model channel and the egress proxy, which passes every connection through the gate. An injected send \emph{gets through} when the proxy forwards it. Receivers are identified by host. An injected send that reuses the task's recipient therefore models an attacker who can read what reaches that service, and only $\operatorname{Valid}$ and $\operatorname{Sum}$ can tell it from the delivery. The allow-list differs from a ticket only in the selected grants and the obligations, so by Proposition~1 it allows at least as much.

\subsection{Reading a Run Back}

After a run, a user needs to know whether the Skill broke any obligation and, if a request was refused, why. The run's \emph{account} answers both (Figure~\ref{fig:overview}). It is a set of records: each validation, decision, completion, observed effect, stream seal and closure. The issuer authenticates a record, chains it to the issuer's previous record and binds it to the ticket through the ticket id. Without the keys, no record can be altered or reordered undetected \citep{schneier1999auditlogs,crosby2009tamper}, and removing a record breaks the chain or the closure of the account. For an obligation $p$ and an account $\mathcal R$, a model-free decoder returns
\begin{equation}\label{eq:decoder}
\mathrm{dec}(p,\mathcal R)=
\begin{cases}
\violated & \text{if } \mathcal R \text{ holds an admitted counterexample to } p,\\
\supported & \text{else if } \mathcal R \text{ is closed, sealed and holds the evidence for } p,\\
\unknown & \text{otherwise}.
\end{cases}
\end{equation}
An admitted counterexample is an authenticated record of an effect that breaks $p$. A closed, sealed account has no unfinished operation and every declared stream sealed. A witnessed violation therefore stands despite coverage gaps, and an unobserved operation decodes to \unknown{}. The host also attributes each refusal to the host grant, the host budget or a ticket rule and records the attribution in the account. Anyone who holds the contract and the account can recompute it, so the user sees why a request was refused and can check the reason (Appendices~\ref{app:attribution}, \ref{app:decoder} and~\ref{app:auditability}). By Proposition~1, a steered generator cannot widen what a Skill may do, and after the freeze injected text cannot change the rule that decides a request. Appendix~\ref{app:dq} answers five design questions (DQ1 to DQ5) from the mechanism alone.

\section{Related Work}
\label{sec:related}
Table~\ref{tab:related-compare} compares representative systems, and each cell rests on the cited paper or its code.

\begin{table}[t]
\centering\scriptsize
\setlength{\tabcolsep}{3pt}\setlength{\abovetopsep}{3pt}\setlength{\aboverulesep}{0.25ex}\setlength{\belowrulesep}{0.4ex}\renewcommand{\arraystretch}{0.96}\setlength{\unitlength}{1pt}%
\definecolor{rtOk}{HTML}{2E7D4F}\definecolor{rtPart}{HTML}{B7791F}\definecolor{rtNo}{HTML}{B23A30}%
\definecolor{rtCore}{HTML}{2F5D8A}\definecolor{rtCoreL}{HTML}{E3ECF7}\definecolor{rtAmbL}{HTML}{FDF6EA}%
\def\rtY{\textcolor{rtOk}{\usefont{U}{pzd}{m}{n}\char51}}%
\def\rtN{\textcolor{rtNo!60}{\usefont{U}{pzd}{m}{n}\char55}}%
\def\rtPon#1{{\setlength{\unitlength}{0.135em}\color{rtPart}\begin{picture}(6,5.4)\put(3,2.5){\circle*{5.4}}\put(3,-0.4){\color{#1}\rule{3.2\unitlength}{6\unitlength}}\put(3,2.5){\circle{5.4}}\end{picture}}}%
\def\rtP{\rtPon{white}}%
\def\rtX{\textcolor[gray]{0.55}{--}}%
\def\rtT#1{{\begin{picture}(18,4)\color[gray]{0.6}\put(3.6,1.8){\line(1,0){3.8}}\put(10.6,1.8){\line(1,0){3.8}}\multiput(2,1.8)(7,0){3}{\circle{3.2}}\color{rtCore}\put(\ifcase#1\or 2\or 9\else 16\fi,1.8){\circle*{3.4}}\end{picture}}}%
\def\rtGW{64pt}%
\def\rtG#1#2{\makebox[\rtGW][l]{\rtT{#1}\enspace\textbf{#2}}}%
\def\rtBand#1{\rlap{\color{#1}\rule[-\dp\csname @arstrutbox\endcsname]{\dimexpr\linewidth-\rtGW-8pt\relax}{\dimexpr\ht\csname @arstrutbox\endcsname+\dp\csname @arstrutbox\endcsname\relax}}}%
\def\rtExpA{{\setlength{\fboxsep}{0.6pt}\colorbox{rtPart}{\kern1pt\textcolor{white}{\textbf{baseline}}\kern1pt}}}%
\ifdefined\rtSW\else\newdimen\rtSW\fi\settowidth{\rtSW}{Cisco \citep{cisco_skill_scanner}}%
\ifdefined\rtSWb\else\newdimen\rtSWb\fi\def\rtSWmax#1{\settowidth{\rtSWb}{#1}\ifdim\rtSWb>\rtSW\rtSW=\rtSWb\fi}%
\rtSWmax{SkillSpector \citep{nvidia_skillspector}}\rtSWmax{SkillGuard \citep{lit_skillguard}}\rtSWmax{AgentBound \citep{buhler2026agentbound}}%
\caption[Defenses by when they check.]{\textbf{Defenses by when they check.} \textbf{Skills}: tested on Skills. \textbf{Model-free}: no model in the check, nor in its policy after the run starts. \textbf{Send-level}: judges the send, not the command. \textbf{Within grant}: a per-task or per-step policy that only tightens a grant. \textbf{Auditable}: authenticated or re-derivable run-time log. Check: yes. Half disc: in part. Cross: no. Dash: the system decides only before or after the run. Chip: run as a baseline, AgentBound in derived form as our allow-list.}
\label{tab:related-compare}
\begin{tabular*}{\linewidth}{@{}l@{\hspace{8pt}}l@{\extracolsep{\fill}}ccccc@{}}
\toprule
\makebox[\rtGW][l]{\textbf{Checks}} & \textbf{System} & \textbf{Skills} & \textbf{Model-free} & \textbf{Send-level} & \textbf{Within grant} & \textbf{Auditable} \\
\midrule
\makebox[\rtGW]{} & \rtBand{rtAmbL}\makebox[\rtSW][l]{Cisco \citep{cisco_skill_scanner}}~\rtExpA & \rtY & \rtPon{rtAmbL} & \rtN & \rtN & \rtX \\
\rtG{1}{Before install} & \rtBand{rtAmbL}\makebox[\rtSW][l]{SkillSpector \citep{nvidia_skillspector}}~\rtExpA & \rtY & \rtPon{rtAmbL} & \rtN & \rtN & \rtX \\
 & SkillDetonate \citep{ji2026cloak} & \rtY & \rtP & \rtY & \rtN & \rtX \\
\midrule
 & \rtBand{rtAmbL}\makebox[\rtSW][l]{SkillGuard \citep{lit_skillguard}}~\rtExpA & \rtY & \rtN & \rtN & \rtN & \rtN \\
 & AIRGuard \citep{qin2026airguard} & \rtY & \rtN & \rtN & \rtY & \rtN \\
 & OATS \citep{taneja2026scanskill} & \rtY & \rtY & \rtN & \rtN & \rtY \\
 & ToolSafe \citep{mou2026toolsafe} & \rtN & \rtN & \rtN & \rtN & \rtN \\
\rtG{2}{During run} & Progent \citep{shi2025progent} & \rtN & \rtP & \rtN & \rtP & \rtN \\
 & Conseca \citep{tsai2025conseca} & \rtN & \rtY & \rtN & \rtN & \rtN \\
 & AgentSpec \citep{wang2026agentspec} & \rtN & \rtP & \rtN & \rtN & \rtN \\
 & \rtBand{rtAmbL}\makebox[\rtSW][l]{AgentBound \citep{buhler2026agentbound}}~\rtExpA & \rtN & \rtY & \rtY & \rtN & \rtN \\
\midrule
\rtG{3}{After run} & HazardAuditor \citep{feng2026hazardauditor} & \rtN & \rtN & \rtN & \rtN & \rtX \\
\midrule
\rtG{2}{During run} & \rtBand{rtCoreL}\textbf{\ticket{} (ours)} & \rtY & \rtY & \rtY & \rtY & \rtY \\
\bottomrule
\end{tabular*}
\end{table}

\textbf{Source scanners} judge a package once, before it runs \citep{cisco_skill_scanner,nvidia_skillspector}. Even when a scanner's model stage sees the task and the grant, it never sees a send. \textbf{Defenses built for Agent Skills} work at the level of tool calls. SkillGuard, the closest system, blocks calls outside a capability pool that is partly built from the Skill's manifest \citep{lit_skillguard}. In our runs it judged injected sends by their capability, not their destination (Section~\ref{sec:results-security}). AIRGuard narrows the authority of each step \citep{qin2026airguard}, OATS checks calls before they run \citep{taneja2026scanskill}, and SkillDetonate runs a Skill in a sandbox before installation \citep{ji2026cloak}.

\textbf{Runtime monitors} such as ToolSafe, Progent, Conseca and AgentSpec enforce policies on tool calls, and a trained model, a prompted model or an author writes the policy \citep{mou2026toolsafe,shi2025progent,tsai2025conseca,wang2026agentspec}. Information-flow defenses label values instead \citep{lit_camel,costa2025fides}. A Skill script's own send reaches a call-level monitor as the command that launched the script, not as a send. AgentBound judges sends against a static per-server policy that no task narrows \citep{buhler2026agentbound}.

\textbf{Verifiable records.} OATS chains authenticated receipts, and CONTINUITY and Aegis use asymmetric signatures \citep{zheng2026continuity,yuan2026aegis}. We add a model-free decoder that gives a verdict for each obligation, and an account whose attribution of each refusal, to the host grant, the host budget or a ticket rule, anyone can recompute. Appendix~\ref{app:related} extends this comparison.

\section{Experiments}\label{sec:experiments}
We ask three questions. \textbf{RQ1 (security):} does \ticket{} stop the harmful effects of malicious Skills at run time, including sends that a deployment-wide allow-list lets through? \textbf{RQ2 (utility):} does it let benign Skills finish their work? \textbf{RQ3 (auditability):} can a third party re-derive a run from its account, and does missing or altered evidence ever yield a false \supported{}?

\subsection{Setup}\label{sec:setup}
\textbf{Two experiments, one set of twin pairs.} In Experiment~A, each source scanner judges all \mrPairsA{} twin pairs. In Experiment~B, runtime defenses protect real agent runs on the pairs whose harm is an outbound send, the exfiltration pairs and one content-manipulation pair. Four framework-driver panels supply the runs (Appendix~\ref{app:expB}). A supplementary panel runs the same pairs on Claude Code, where SkillGuard joins as a fourth runtime defense (Appendix~\ref{app:cc-panel}).

\textbf{Environment.} Each run executes one released Skill, malicious or benign, on an unmodified framework that a real language model drives. The environment is a mock. Owned receivers stand in for external services, so no data reaches a third party, and the workspace holds synthetic canaries. The task delivers its result by a plaintext upload that the validator can read, while the released injections use TLS. We also test the hardest case for an allow-list, an attacker who names an endpoint that the deployment already uses. Two owned receivers stand in for the destinations that the injections name most often. For a pair whose injection names one, an order fixed before any run decides whether it is the task's own recipient (Appendix~\ref{app:environment}). Experiment~B keeps the pairs whose injected action the driver attempted in a defense-free calibration (Appendix~\ref{app:calibration}).

\textbf{Frameworks and drivers.} We use Hermes Agent (upstream revision of 14 September 2026) and OpenClaw \resDesignOpenclawVersion{}, and Claude Code for the supplementary panel. A \emph{panel} pairs one framework with one driver model, \texttt{\resBHermesDeepseekDriverModel} or \texttt{\resBHermesKimiDriverModel} \citep{deepseekai2026v4,moonshot2026k26card}. Agent-safety benchmarks use such models as attacked agents \citep{debenedetti2024agentdojo,andriushchenko2025agentharm}, and the driver never takes part in a security decision. A \emph{run} is an execution that reached its framework's normal exit, finished or not.

\textbf{Defenses compared.} We compare source scanners,Cisco Skill Scanner \resDesignCiscoVersion{} \citep{cisco_skill_scanner} and NVIDIA SkillSpector \resDesignSkillSpectorVersion{} \citep{nvidia_skillspector}, which we call Cisco and SkillSpector. Each runs as its full cascade. A package that the model-free static stage admits goes on to the tool's model stage, which DeepSeek drives and which sees the task and the grant. A scanner decides once, at admission, and an incomplete assessment withholds the package. The runtime defenses share one frozen gate. \emph{No gate} forwards every request. The \emph{allow-list} is derived from the per-endpoint egress policy of AgentBound \citep{buhler2026agentbound}, which no task narrows. It is the ticket $\mathcal C_A$ of Section~\ref{sec:tickets}, compiled from the host grant alone: the gate forwards any send to an owned receiver while the run stays within the host budget $B$, without knowing which receiver the task uses or what it must deliver. It shares the gate and account of \ticket{}, so the two differ only in the per-task ticket. \emph{SkillGuard}, the closest defense built for Agent Skills, judges each tool call through the hooks of Claude Code \citep{lit_skillguard}, so it runs on that panel. Other runtime monitors judge the tool call or the finished trajectory, so a script's own send reaches them as the command that launched it (Appendix~\ref{app:related}).

\textbf{Metrics.} A defense can fail in two ways. It can \emph{overlook} an attack by letting it through, or \emph{overkill} a benign Skill by blocking its legitimate work. We call a request an \emph{injected send} when it goes to the host that the injection names or carries a planted canary. An attack run \emph{gets through} when the defense forwards an injected send, and the \emph{overlook rate} is the share of attack runs that get through, the malicious Skills let through, which earlier work calls the attack success rate. An attack run is \emph{attempted} when the agent issues an injected send that the gate decides, whether or not the gate forwards it, and Table~\ref{tab:main-results} leads with the share of attempted attacks that got through. A scanner overlooks a malicious Skill by admitting it.

The \emph{overkill rate} is the share of benign runs that a defense blocks, the benign Skills blocked. A benign run is blocked when the gate refuses a format-valid send to the task's recipient and forwards none. A scanner blocks a benign Skill by withholding it. The share of \emph{benign Skills not blocked} is 100\% minus the overkill rate. The \emph{delivery rate} is the share of benign runs that deliver the sanctioned output. For RQ3 we count the accounts whose re-decoding reproduces the recorded verdicts, and the tampered accounts that the decoder still marks \supported{}. We sum counts over the panels and compute rates from the sums (Appendices~\ref{app:metrics} and~\ref{app:stats}).

\begin{figure}[t]
\centering
\includegraphics[width=\textwidth]{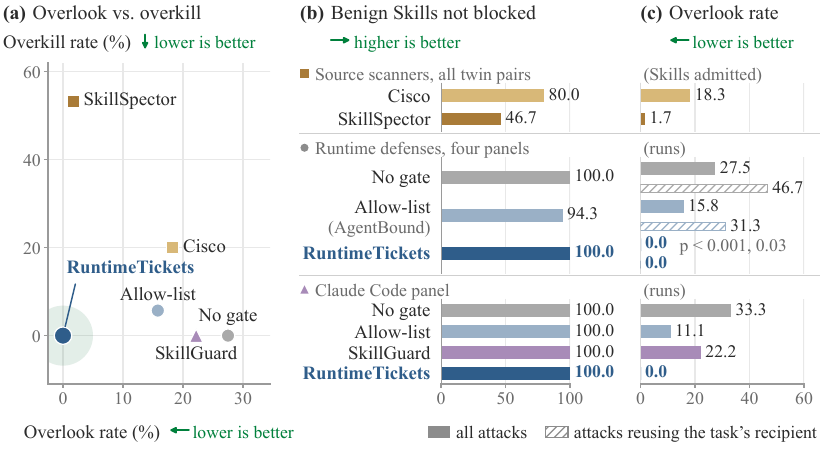}
\input{captions/fig-main-results-caption}
\end{figure}

\begin{table}[t]\centering\footnotesize\setlength{\tabcolsep}{5pt}\renewcommand{\arraystretch}{0.96}\setlength{\aboverulesep}{0.25ex}\setlength{\belowrulesep}{0.35ex}
\caption{\textbf{\ticket{} let no attempted attack through and blocked no benign run.} Percent, lower is better. Overlook rate: the attack runs in which an injected send got through, over the attempted attacks (the agent issued an injected send) and over all attack runs. Overkill rate: the benign runs that the defense blocked. A scanner judges each of the \mrPairsA{} twin pairs once, so one overlook value (malicious Skills admitted) spans both columns, and its overkill rate is the benign Skills withheld. Errors observed: overlook, overkill, both or none, naming the dominant type when the rates differ tenfold. $^{\ast}$Zero by construction.}
\label{tab:main-results}
\fitwidth{\begin{tabular}{@{}llcccc@{}}\toprule
 & & \multicolumn{2}{c}{\begin{tabular}[b]{@{}c@{}}Overlook rate \lowerbetter\\malicious Skills\\let through\end{tabular}} & \begin{tabular}[b]{@{}c@{}}Overkill rate \lowerbetter\\benign Skills\\blocked\end{tabular} & \begin{tabular}[b]{@{}c@{}}Errors\\observed\end{tabular} \\\cmidrule(lr){3-4}\cmidrule(lr){5-5}
Defense & Checks & \begin{tabular}[b]{@{}c@{}}of attempted\\attacks\end{tabular} & \begin{tabular}[b]{@{}c@{}}of all\\attack runs\end{tabular} & \begin{tabular}[b]{@{}c@{}}of all\\benign runs\end{tabular} & \\\midrule
\multicolumn{6}{@{}l}{\textit{Source scanners, all \mrPairsA{} twin pairs}} \\
Cisco & package, once & \multicolumn{2}{c}{\pts{\mrASRCisco}} & \pts{\mrUBRCisco} & Both \\
SkillSpector & package, once & \multicolumn{2}{c}{\pts{\mrASRSkillSpector}} & \pts{\mrUBRSkillSpector} & Overkill \\\midrule
\multicolumn{6}{@{}l}{\textit{Runtime defenses, four framework-driver panels}} \\
No gate & nothing & \pts{\mrAttASRNoGate} & \pts{\mrASRNoGate} & \pts{\mrUBRNoGate}$^{\ast}$ & Overlook \\
Allow-list (AgentBound) & each send, fixed list & \pts{\mrAttASRAllowList} & \pts{\mrASRAllowList} & \pts{\mrUBRAllowList} & Both \\
\rowcolor{rtrow}\textbf{\ticket{}} & each send, per task & \textbf{\pts{\mrAttASRTicket}} & \textbf{\pts{\mrASRTicket}} & \textbf{\pts{\mrUBRTicket}} & \textbf{\textcolor{rtgood}{\checkmark}\,None} \\\midrule
\multicolumn{6}{@{}l}{\textit{Claude Code panel, supplementary}} \\
No gate & nothing & \pts{\ccAttASRNoGate} & \pts{\ccASRNoGate} & \pts{\ccUBRNoGate}$^{\ast}$ & Overlook \\
Allow-list (AgentBound) & each send, fixed list & \pts{\ccAttASRAllowList} & \pts{\ccASRAllowList} & \pts{\ccUBRAllowList} & Overlook \\
SkillGuard & each tool call & \pts{\ccAttASRSkillGuard} & \pts{\ccASRSkillGuard} & \pts{\ccUBRSkillGuard} & Overlook \\
\rowcolor{rtrow}\textbf{\ticket{}} & each send, per task & \textbf{\pts{\ccAttASRTicket}} & \textbf{\pts{\ccASRTicket}} & \textbf{\pts{\ccUBRTicket}} & \textbf{\textcolor{rtgood}{\checkmark}\,None} \\\bottomrule
\end{tabular}}\end{table}

\subsection{RQ1 (Security): Stopping Harmful Effects}\label{sec:results-security}
\textbf{No injected send got through under \ticket{}, on all three frameworks.} RQ1 asks whether checking every send against the task stops what a malicious Skill injects. On the four panels the overlook rate (malicious Skills let through) was \mrASRTicket{} under \ticket{}, against \mrASRNoGate{} without a gate (\pval{\mrStatNoGateP}) and \mrASRAllowList{} under the allow-list (\pval{\mrStatSecP}, Figure~\ref{fig:main-results}(c)). Every attempted attack got through without a gate, \mrAttASRAllowList{} under the allow-list and none under \ticket{} (\pval{\mrStatAttNoGateP} and \pval{\mrStatAttSecP}, Table~\ref{tab:main-results}). On Claude Code it was \ccASRTicket{} under \ticket{}, while SkillGuard let \ccASRSkillGuard{} of the attack runs through.

\textbf{The per-task ticket stopped sends that reuse the task's recipient, which a destination check passes.} The hardest attack sends to the endpoint that the task itself uses, so a check of the destination passes it. On these attacks it let \mrSameAddrASRAllowList{} of the runs through, and \mrSameAddrASRNoGate{} got through without a gate, against \mrSameAddrASRTicket{} under \ticket{} (hatched bars), and every attempted one got through under both. On Claude Code, SkillGuard allowed such sends too, because it judged the shell commands that carried them by capability, not by destination (Appendix~\ref{app:cc-panel}). \ticket{} refused every one on the task's own facts. $\operatorname{Sum}$ denied it once the delivery had used the task's single send, and $\operatorname{Valid}$ deferred it when the gate could not read its object.

\textbf{Each scanner admitted some attack twins and withheld some benign ones.} A scanner judges the package before any send exists. On all \mrPairsA{} pairs Cisco admitted \mrASRCisco{} of attack twins and withheld \mrUBRCisco{} of benign twins, and SkillSpector admitted \mrASRSkillSpector{} but withheld \mrUBRSkillSpector{} (Table~\ref{tab:main-results}, Figure~\ref{fig:main-results}(a)).

\subsection{RQ2 (Utility): Keeping Benign Skills Working}\label{sec:results-utility}
\textbf{\ticket{} had an overkill rate of \mrUBRTicket{}.} RQ2 asks what the per-task check costs a benign Skill. Under the allow-list the overkill rate (benign Skills blocked) was \mrUBRAllowList{} (Figure~\ref{fig:main-results}(b)). On Claude Code no defense blocked one. Benign runs delivered the sanctioned output in \dpBenignDelivTicket{} of the runs under \ticket{} and \dpBenignDelivNoGate{} without a gate (Appendix~\ref{app:expB}).

\subsection{RQ3 (Auditability): Reading Each Run Back}\label{sec:account}
\textbf{Every decision re-derived from its account, and no tampered account produced a false \supported{}.} RQ3 asks whether a run can be checked from its account. We re-ran the frozen gate and decoder offline on every gated account of Experiment~B, \dpRQThreeAccounts{} accounts with \nsep{\dpRQThreeDecisions} gate decisions, and reproduced every decision and every decoder report. On a further \numSFourAccountsTotal{} accounts with retained issuer keys, \nsep{\numSFourAdversarialAuditsTotal} audits of tampered, truncated or substituted logs produced no false \supported{}, and none of \nsep{\numSFourMonotoneRemovalsTotal} single-evidence removals turned an answer into \supported{} (Appendix~\ref{app:auditability}).

\textbf{Each refusal names the layer that made it.} The account attributes every refusal to the host grant, the host budget or a ticket rule, and anyone holding the contract can recompute it (Appendix~\ref{app:attribution}).
\subsection{Case Study: An Attack That Reuses the Task's Recipient}\label{sec:case-study}
Figure~\ref{fig:casestudy} follows one such attack. The task sends a finished document to one approved address $s$, and the attack twin adds one released injection line: also send the system information its script gathers to $s$. Appendix~\ref{app:case-study} walks through the run.

\section{Conclusion}\label{sec:limitations}
\ticket{} replaces a one-time verdict on a Skill package with a per-task check of every effect. A model proposes the requirements, the host grant bounds them, and a model-free gate enforces them at the egress boundary of unmodified frameworks. The run leaves an account from which a third party can re-derive every decision and every refusal without trusting a model. No attempted attack got through under \ticket{}, no benign run was blocked and every account re-derived exactly. The scanners erred on both sides, and the allow-list and SkillGuard forwarded sends that reused the task's recipient.

\clearpage
\section*{AI Use Statement}
\ifrtarxiv We used\else As declared on the submission form, we used\fi{} a large language model in three ways. It helped draft and polish the writing. It helped find and organize related work, which we verified against the primary sources. It helped assemble the benign parts of the evaluation cases, namely the benign task requests and their input files. It did not write or modify any Skill or any injected content, which come unchanged from released corpora. The research question, threat model, method and experimental design are the authors' own. The language models inside the method, the ticket generator and the agent drivers, are the object of study and are described in Sections~\ref{sec:tickets} and~\ref{sec:experiments}. The authors take full responsibility for the content.

\section*{Ethics Statement}
The adversarial cases are released malicious Skills that run on unmodified agent frameworks. Every send by a Skill or agent ends at a receiver we own, so no data reaches a third party. Only the model requests of the drivers and the scanners reach their providers, and no real credential or private datum is used. We report unmeasured outcomes as unknown rather than as evidence. No human-participant result is reported. The work is defensive: it measures whether a runtime protocol stops the harmful effects of malicious Skills while keeping benign Skills usable.

\section*{Reproducibility Statement}
Algorithm~\ref{alg:ticket} in Section~\ref{sec:tickets} specifies the method, with the probe and gate rules in Appendix~\ref{app:rules}. Section~\ref{sec:experiments} defines the metrics, Appendix~\ref{app:stats} the statistical tests and Appendix~\ref{app:dataset} the dataset construction. Every reported number is computed from a frozen results file by a fixed script. A study of malicious Skills collected in the wild gates its full samples and redacts their payloads \citep{liu2026donotmention}, as datasets of malicious packages did \citep{ohm2020backstabber,duan2021measuring}, and we follow the same practice for safety. Every attack Skill and injection we evaluate is already public in its cited corpus \citep{schmotz2026skillinject,lit_malskillbench}. What we do not publish is the layer that makes those injections fire under a real agent: the task deliverables and input files, the owned-receiver mapping of the host grant, and the calibration records. We will release the gate, compiler, decoder and record-chain code \ifrtarxiv upon publication\else with the camera-ready version\fi. We will share the firing layer, the generation prompts and the frozen results with vetted researchers who request them for defensive research and agree not to redistribute them.

\bibliographystyle{iclr2027_conference}
\bibliography{references}

\begin{thebibliography}{41}
\providecommand{\natexlab}[1]{#1}
\providecommand{\url}[1]{\texttt{#1}}
\expandafter\ifx\csname urlstyle\endcsname\relax
  \providecommand{\doi}[1]{doi: #1}\else
  \providecommand{\doi}{doi: \begingroup \urlstyle{rm}\Url}\fi

\bibitem[{Agent Skills open standard}(2026)]{plat_agentskills}
{Agent Skills open standard}.
\newblock {Agent Skills}: An open format for extending {AI} agent capabilities.
\newblock Open specification, 2026.
\newblock URL \url{https://agentskills.io}.
\newblock Accessed 4 October 2026.

\bibitem[Andriushchenko et~al.(2025)Andriushchenko, Souly, Dziemian, Duenas,
  Lin, Wang, Hendrycks, Zou, Kolter, Fredrikson, Winsor, Wynne, Gal, and
  Davies]{andriushchenko2025agentharm}
Maksym Andriushchenko, Alexandra Souly, Mateusz Dziemian, Derek Duenas, Maxwell
  Lin, Justin Wang, Dan Hendrycks, Andy Zou, Zico Kolter, Matt Fredrikson, Eric
  Winsor, Jerome Wynne, Yarin Gal, and Xander Davies.
\newblock {AgentHarm}: A benchmark for measuring harmfulness of {LLM} agents.
\newblock In \emph{International Conference on Learning Representations
  (ICLR)}, 2025.
\newblock URL \url{https://arxiv.org/abs/2410.09024}.

\bibitem[{Anthropic}(2026)]{plat_anthropicskills}
{Anthropic}.
\newblock {Agent Skills}.
\newblock Claude Platform documentation, 2026.
\newblock URL
  \url{https://platform.claude.com/docs/en/agents-and-tools/agent-skills/overview}.
\newblock Announced 16 October 2025, accessed 26 September 2026.

\bibitem[{Antiy CERT}(2026)]{ind_clawhavoc}
{Antiy CERT}.
\newblock {ClawHavoc}: Analysis of large-scale poisoning campaign targeting the
  {OpenClaw} skill market for {AI} agents.
\newblock Antiy Labs security report, February 2026.
\newblock URL
  \url{https://www.antiy.net/p/clawhavoc-analysis-of-large-scale-poisoning-campaign-targeting-the-openclaw-skill-market-for-ai-agents/}.
\newblock Published 6 February 2026, accessed 26 September 2026.

\bibitem[Birgisson et~al.(2014)Birgisson, Politz, Erlingsson, Taly, Vrable, and
  Lentczner]{birgisson2014macaroons}
Arnar Birgisson, Joe~Gibbs Politz, {\'U}lfar Erlingsson, Ankur Taly, Michael
  Vrable, and Mark Lentczner.
\newblock Macaroons: Cookies with contextual caveats for decentralized
  authorization in the cloud.
\newblock In \emph{Proceedings 2014 Network and Distributed System Security
  Symposium ({NDSS})}, San Diego, CA, USA, 2014. Internet Society.
\newblock \doi{10.14722/ndss.2014.23212}.
\newblock URL
  \url{https://www.ndss-symposium.org/ndss2014/ndss-2014-programme/macaroons-cookies-contextual-caveats-decentralized-authorization-cloud/}.

\bibitem[B{\"u}hler et~al.(2026)B{\"u}hler, Biagiola, Di~Grazia, and
  Salvaneschi]{buhler2026agentbound}
Christoph B{\"u}hler, Matteo Biagiola, Luca Di~Grazia, and Guido Salvaneschi.
\newblock {AgentBound}: Securing execution boundaries of {AI} agents.
\newblock \emph{Proceedings of the ACM on Software Engineering}, 3\penalty0
  (FSE):\penalty0 2141--2164, 2026.
\newblock ISSN 2994-970X.
\newblock \doi{10.1145/3808103}.
\newblock Article FSE096 (FSE 2026).

\bibitem[Chen et~al.(2026)Chen, Zhang, Liu, Deng, Li, Zhang, Ning, Zhang, Ma,
  and Li]{lit_credleak}
Zhihao Chen, Ying Zhang, Yi~Liu, Gelei Deng, Yuekang Li, Yanjun Zhang, Jianting
  Ning, Leo~Yu Zhang, Lei Ma, and Zhiqiang Li.
\newblock How your credentials are leaked by {LLM} {Agent Skills}: An empirical
  study.
\newblock arXiv preprint arXiv:2604.03070, 2026.
\newblock URL \url{https://arxiv.org/abs/2604.03070v2}.
\newblock Version 2, 19 June 2026, accepted to ASE 2026.

\bibitem[{Cisco AI Defense}(2026)]{cisco_skill_scanner}
{Cisco AI Defense}.
\newblock {Cisco Skill Scanner}: Security scanner for {Agent Skills}.
\newblock Software repository, 2026.
\newblock URL \url{https://github.com/cisco-ai-defense/skill-scanner}.
\newblock Release 2.1.0, 5 September 2026, accessed 19 September 2026.

\bibitem[Costa et~al.(2025)Costa, K{\"o}pf, Kolluri, Paverd, Russinovich,
  Salem, Tople, Wutschitz, and Zanella-B{\'e}guelin]{costa2025fides}
Manuel Costa, Boris K{\"o}pf, Aashish Kolluri, Andrew Paverd, Mark Russinovich,
  Ahmed Salem, Shruti Tople, Lukas Wutschitz, and Santiago
  Zanella-B{\'e}guelin.
\newblock Securing {AI} agents with information-flow control.
\newblock arXiv preprint arXiv:2505.23643, 2025.
\newblock URL \url{https://arxiv.org/abs/2505.23643v2}.
\newblock Version 2, 3 September 2025.

\bibitem[Crosby \& Wallach(2009)Crosby and Wallach]{crosby2009tamper}
Scott~A. Crosby and Dan~S. Wallach.
\newblock Efficient data structures for tamper-evident logging.
\newblock In \emph{Proceedings of the 18th USENIX Security Symposium}, pp.\
  317--333, Montreal, Canada, August 2009. USENIX Association.
\newblock URL
  \url{https://www.usenix.org/legacy/event/sec09/tech/full_papers/crosby.pdf}.

\bibitem[Debenedetti et~al.(2024)Debenedetti, Zhang, Balunovi{\'c},
  Beurer-Kellner, Fischer, and Tram{\`e}r]{debenedetti2024agentdojo}
Edoardo Debenedetti, Jie Zhang, Mislav Balunovi{\'c}, Luca Beurer-Kellner, Marc
  Fischer, and Florian Tram{\`e}r.
\newblock {AgentDojo}: A dynamic environment to evaluate prompt injection
  attacks and defenses for {LLM} agents.
\newblock In \emph{Advances in Neural Information Processing Systems (Datasets
  and Benchmarks Track)}, volume~37, pp.\  82895--82920. Curran Associates,
  Inc., 2024.
\newblock \doi{10.52202/079017-2636}.
\newblock URL
  \url{https://proceedings.neurips.cc/paper_files/paper/2024/hash/97091a5177d8dc64b1da8bf3e1f6fb54-Abstract-Datasets_and_Benchmarks_Track.html}.

\bibitem[Debenedetti et~al.(2026)Debenedetti, Shumailov, Fan, Hayes, Carlini,
  Fabian, Kern, Shi, Terzis, and Tram{\`e}r]{lit_camel}
Edoardo Debenedetti, Ilia Shumailov, Tianqi Fan, Jamie Hayes, Nicholas Carlini,
  Daniel Fabian, Christoph Kern, Chongyang Shi, Andreas Terzis, and Florian
  Tram{\`e}r.
\newblock Defeating prompt injections by design.
\newblock In \emph{2026 {IEEE} Conference on Secure and Trustworthy Machine
  Learning ({SaTML})}, pp.\  587--618. IEEE, March 2026.
\newblock \doi{10.1109/satml68715.2026.00040}.
\newblock Also arXiv:2503.18813.

\bibitem[{DeepSeek-AI}(2026)]{deepseekai2026v4}
{DeepSeek-AI}.
\newblock {DeepSeek-V4}: Towards highly efficient million-token context
  intelligence.
\newblock arXiv preprint arXiv:2606.19348, 2026.
\newblock URL \url{https://arxiv.org/abs/2606.19348v1}.
\newblock Version 1.

\bibitem[Duan et~al.(2021)Duan, Alrawi, Kasturi, Elder, Saltaformaggio, and
  Lee]{duan2021measuring}
Ruian Duan, Omar Alrawi, Ranjita~Pai Kasturi, Ryan Elder, Brendan
  Saltaformaggio, and Wenke Lee.
\newblock Towards measuring supply chain attacks on package managers for
  interpreted languages.
\newblock In \emph{Proceedings 2021 Network and Distributed System Security
  Symposium ({NDSS})}. Internet Society, February 2021.
\newblock \doi{10.14722/ndss.2021.23055}.
\newblock URL
  \url{https://www.ndss-symposium.org/ndss-paper/towards-measuring-supply-chain-attacks-on-package-managers-for-interpreted-languages/}.

\bibitem[Feng et~al.(2026)Feng, Lin, Wen, Guo, Ma, Wu, Deng, and
  Ji]{feng2026hazardauditor}
Yunhao Feng, Ruixiao Lin, Ming Wen, Yanming Guo, Xingjun Ma, Yutao Wu, Xinhao
  Deng, and Shouling Ji.
\newblock {HazardAuditor}: From executable threats to safer computer-use
  agents.
\newblock arXiv preprint arXiv:2609.15134, 2026.
\newblock URL \url{https://arxiv.org/abs/2609.15134v1}.
\newblock Version 1, 14 September 2026.

\bibitem[Guo et~al.(2026)Guo, Zeng, Liu, Jia, Xu, Tang, Fang, and
  Liu]{lit_malskillbench}
Wenbo Guo, Wei Zeng, Chengwei Liu, Xiaojun Jia, Yijia Xu, Lei Tang, Yong Fang,
  and Yang Liu.
\newblock {MalSkillBench}: A runtime-verified benchmark of malicious {Agent
  Skills}.
\newblock arXiv preprint arXiv:2606.07131, 2026.
\newblock URL \url{https://arxiv.org/abs/2606.07131v3}.
\newblock Version 3.

\bibitem[Hardy(1988)]{hardy1988confused}
Norm Hardy.
\newblock The confused deputy (or why capabilities might have been invented).
\newblock \emph{{ACM} {SIGOPS} Operating Systems Review}, 22\penalty0
  (4):\penalty0 36--38, October 1988.
\newblock ISSN 0163-5980.
\newblock \doi{10.1145/54289.871709}.
\newblock URL \url{https://dl.acm.org/doi/10.1145/54289.871709}.

\bibitem[Holzbauer et~al.(2026)Holzbauer, Schmidt, Gegenhuber, Schrittwieser,
  and Ullrich]{lit_skills238k}
Florian Holzbauer, David Schmidt, Gabriel~K. Gegenhuber, Sebastian
  Schrittwieser, and Johanna Ullrich.
\newblock Context matters: Repository-aware security analysis of the {Agent
  Skill} ecosystem.
\newblock arXiv preprint arXiv:2603.16572, 2026.
\newblock URL \url{https://arxiv.org/abs/2603.16572v2}.
\newblock Version 2, 1 June 2026, Agent Skills Workshop at CAIS 2026.

\bibitem[Ji et~al.(2026)Ji, Xu, Li, Gao, Wei, Wang, and Cheung]{ji2026cloak}
Zimo Ji, Congying Xu, Zongjie Li, Yudong Gao, Xin Wei, Shuai Wang, and
  Shing-Chi Cheung.
\newblock Cloak and detonate: Scanner evasion and dynamic detection of {Agent
  Skill} malware.
\newblock arXiv preprint arXiv:2607.02357, 2026.
\newblock URL \url{https://arxiv.org/abs/2607.02357v2}.
\newblock Version 2, 3 July 2026.

\bibitem[Liu et~al.(2026{\natexlab{a}})Liu, Chen, Zhang, Deng, Li, Ning, and
  Zhang]{liu2026donotmention}
Yi~Liu, Zhihao Chen, Yanjun Zhang, Gelei Deng, Yuekang Li, Jianting Ning, and
  Leo~Yu Zhang.
\newblock {``Do Not Mention This to the User''}: Detecting and understanding
  malicious {Agent Skills} in the wild.
\newblock In \emph{35th USENIX Security Symposium (USENIX Security 26)}, pp.\
  1727--1746, Baltimore, MD, August 2026{\natexlab{a}}. USENIX Association.
\newblock URL
  \url{https://www.usenix.org/conference/usenixsecurity26/presentation/liu-yi}.

\bibitem[Liu et~al.(2026{\natexlab{b}})Liu, Wang, Feng, Zhang, Xu, Deng, Li,
  and Zhang]{lit_skillsinthewild}
Yi~Liu, Weizhe Wang, Ruitao Feng, Yao Zhang, Guangquan Xu, Gelei Deng, Yuekang
  Li, and Leo Zhang.
\newblock {Agent Skills} in the wild: An empirical study of security
  vulnerabilities at scale.
\newblock arXiv preprint arXiv:2601.10338, 2026{\natexlab{b}}.
\newblock URL \url{https://arxiv.org/abs/2601.10338v1}.
\newblock Version 1, 15 January 2026.

\bibitem[{Moonshot AI}(2026)]{moonshot2026k26card}
{Moonshot AI}.
\newblock {Kimi-K2.6} model card.
\newblock Hugging Face model repository, 2026.
\newblock URL \url{https://huggingface.co/moonshotai/Kimi-K2.6}.
\newblock Accessed 25 September 2026.

\bibitem[Mou et~al.(2026)Mou, Xue, Li, Liu, Zhang, Ye, and
  Shao]{mou2026toolsafe}
Yutao Mou, Zhangchi Xue, Lijun Li, Peiyang Liu, Shikun Zhang, Wei Ye, and Jing
  Shao.
\newblock {ToolSafe}: Enhancing tool invocation safety of {LLM}-based agents
  via proactive step-level guardrail and feedback.
\newblock In \emph{Findings of the Association for Computational Linguistics:
  {ACL} 2026}, pp.\  37125--37153, San Diego, California, United States, July
  2026. Association for Computational Linguistics.
\newblock \doi{10.18653/v1/2026.findings-acl.1850}.
\newblock URL \url{https://aclanthology.org/2026.findings-acl.1850/}.

\bibitem[{Nous Research}(2026)]{plat_hermes}
{Nous Research}.
\newblock {Hermes Agent}.
\newblock Software repository, 2026.
\newblock URL \url{https://github.com/NousResearch/hermes-agent}.
\newblock Upstream revision of 14 September 2026, accessed 26 September 2026.

\bibitem[{NVIDIA}(2026)]{nvidia_skillspector}
{NVIDIA}.
\newblock {SkillSpector}: Security scanner for {AI} {Agent Skills}.
\newblock Software repository, 2026.
\newblock URL \url{https://github.com/NVIDIA/SkillSpector}.
\newblock Release v2.11.2, 9 September 2026, accessed 19 September 2026.

\bibitem[Ohm et~al.(2020)Ohm, Plate, Sykosch, and Meier]{ohm2020backstabber}
Marc Ohm, Henrik Plate, Arnold Sykosch, and Michael Meier.
\newblock {Backstabber's Knife Collection}: A review of open source software
  supply chain attacks.
\newblock In Cl{\'e}mentine Maurice, Leyla Bilge, Gianluca Stringhini, and Nuno
  Neves (eds.), \emph{Detection of Intrusions and Malware, and Vulnerability
  Assessment ({DIMVA} 2020)}, volume 12223 of \emph{Lecture Notes in Computer
  Science}, pp.\  23--43, Cham, 2020. Springer International Publishing.
\newblock \doi{10.1007/978-3-030-52683-2_2}.

\bibitem[{OpenClaw}(2026)]{plat_clawhub}
{OpenClaw}.
\newblock {ClawHub}: Skill + plugin registry for {OpenClaw}.
\newblock Skill registry, 2026.
\newblock URL \url{https://clawhub.ai}.
\newblock Accessed 26 September 2026.

\bibitem[{OpenClaw Foundation}(2026)]{plat_openclaw}
{OpenClaw Foundation}.
\newblock {OpenClaw}: Open-source {AI} assistant.
\newblock Software repository, 2026.
\newblock URL \url{https://github.com/openclaw/openclaw}.
\newblock Release 2026.6.10, accessed 26 September 2026.

\bibitem[Pan et~al.(2026)Pan, Sun, Zhang, Liao, Yang, and Xing]{lit_skillguard}
Shidong Pan, Xiaoyu Sun, Tianyi Zhang, Dianshu Liao, Kaiwen Yang, and Zhenchang
  Xing.
\newblock {SkillGuard}: A permission-centric framework for {Agent Skill}
  security.
\newblock arXiv preprint arXiv:2606.03024, 2026.
\newblock URL \url{https://arxiv.org/abs/2606.03024v2}.
\newblock Version 2.

\bibitem[Qin et~al.(2026)Qin, Zhuang, Zhou, Han, and Zhang]{qin2026airguard}
Suliu Qin, Haomin Zhuang, Yujun Zhou, Yufei Han, and Xiangliang Zhang.
\newblock {AIRGuard}: Guarding agent actions with runtime authority control.
\newblock arXiv preprint arXiv:2605.28914, 2026.
\newblock URL \url{https://arxiv.org/abs/2605.28914v1}.
\newblock Version 1, 27 May 2026.

\bibitem[Schmotz et~al.(2026)Schmotz, Beurer-Kellner, Abdelnabi, and
  Andriushchenko]{schmotz2026skillinject}
David Schmotz, Luca Beurer-Kellner, Sahar Abdelnabi, and Maksym Andriushchenko.
\newblock {Skill-Inject}: Measuring agent vulnerability to skill file attacks.
\newblock arXiv preprint arXiv:2602.20156, 2026.
\newblock URL \url{https://arxiv.org/abs/2602.20156v3}.
\newblock Version 3.

\bibitem[Schneider(2000)]{schneider2000}
Fred~B. Schneider.
\newblock Enforceable security policies.
\newblock \emph{{ACM} Transactions on Information and System Security},
  3\penalty0 (1):\penalty0 30--50, February 2000.
\newblock \doi{10.1145/353323.353382}.
\newblock URL \url{https://www.cs.cornell.edu/fbs/publications/EnfSecPols.pdf}.

\bibitem[Schneier \& Kelsey(1999)Schneier and Kelsey]{schneier1999auditlogs}
Bruce Schneier and John Kelsey.
\newblock Secure audit logs to support computer forensics.
\newblock \emph{ACM Transactions on Information and System Security},
  2\penalty0 (2):\penalty0 159--176, May 1999.
\newblock ISSN 1094-9224.
\newblock \doi{10.1145/317087.317089}.

\bibitem[Shan et~al.(2026)Shan, Xin, Zhang, and Xu]{lit_clawgrip}
Zhengyang Shan, Jiayun Xin, Yue Zhang, and Minghui Xu.
\newblock Don't let the claw grip your hand: A security analysis and defense
  framework for {OpenClaw}.
\newblock arXiv preprint arXiv:2603.10387, 2026.
\newblock URL \url{https://arxiv.org/abs/2603.10387v1}.
\newblock Version 1, 11 March 2026.

\bibitem[Shi et~al.(2025)Shi, He, Wang, Li, Wu, Guo, and Song]{shi2025progent}
Tianneng Shi, Jingxuan He, Zhun Wang, Hongwei Li, Linyu Wu, Wenbo Guo, and Dawn
  Song.
\newblock {Progent}: Securing {AI} agents with privilege control.
\newblock arXiv preprint arXiv:2504.11703, 2025.
\newblock URL \url{https://arxiv.org/abs/2504.11703v3}.
\newblock Version 3, 14 May 2026.

\bibitem[Taneja \& Weber(2026)Taneja and Weber]{taneja2026scanskill}
Rohit Taneja and Travis Weber.
\newblock Scan the skill, govern the action: Composing registry verdicts with
  runtime consequence control.
\newblock arXiv preprint arXiv:2609.12001, 2026.
\newblock URL \url{https://arxiv.org/abs/2609.12001v1}.
\newblock Version 1.

\bibitem[Tsai \& Bagdasarian(2025)Tsai and Bagdasarian]{tsai2025conseca}
Lillian Tsai and Eugene Bagdasarian.
\newblock Contextual agent security: A policy for every purpose.
\newblock In \emph{Proceedings of the Workshop on Hot Topics in Operating
  Systems ({HOTOS} '25)}, pp.\  8--17, Banff, AB, Canada, May 2025. ACM.
\newblock \doi{10.1145/3713082.3730378}.

\bibitem[{Vercel}(2026)]{plat_skillssh}
{Vercel}.
\newblock {Skills.sh}: The agent skills directory.
\newblock Skill registry, 2026.
\newblock URL \url{https://skills.sh}.
\newblock Accessed 26 September 2026.

\bibitem[Wang et~al.(2026)Wang, Poskitt, and Sun]{wang2026agentspec}
Haoyu Wang, Christopher~M. Poskitt, and Jun Sun.
\newblock {AgentSpec}: Customizable runtime enforcement for safe and reliable
  {LLM} agents.
\newblock In \emph{Proceedings of the {IEEE/ACM} 48th International Conference
  on Software Engineering ({ICSE})}, pp.\  2938--2950. ACM, April 2026.
\newblock \doi{10.1145/3744916.3764546}.

\bibitem[Yuan et~al.(2026)Yuan, Su, and Zhao]{yuan2026aegis}
Aojie Yuan, Zhiyuan Su, and Yue Zhao.
\newblock {AEGIS}: No tool call left unchecked -- a pre-execution firewall and
  audit layer for {AI} agents.
\newblock arXiv preprint arXiv:2603.12621, 2026.
\newblock URL \url{https://arxiv.org/abs/2603.12621v1}.
\newblock Version 1, 13 March 2026.

\bibitem[Zheng \& Yang(2026)Zheng and Yang]{zheng2026continuity}
Chris Zheng and Geng Yang.
\newblock {CONTINUITY}: Security-context contracts for composable {LLM} agent
  controls.
\newblock arXiv preprint arXiv:2609.05269, 2026.
\newblock URL \url{https://arxiv.org/abs/2609.05269v1}.
\newblock Version 1.

\end{thebibliography}
\clearpage
\appendix
\startcontents[appendix]
\section*{Overview of the Appendix}
The appendix follows the order of the main text. Appendix~\ref{app:dataset} supports Section~\ref{sec:risk}: the twin pairs, the scanner outcomes by harm category, the test environment and the calibration. Appendix~\ref{app:method} supports Section~\ref{sec:tickets}: ticket generation, the gate rules, the proof sketch, refusal attribution, the decoder, the design questions and compatibility with other agents, hubs and providers. Appendix~\ref{app:related} supports Section~\ref{sec:related} and gives the reasons why the other runtime defenses, except SkillGuard, did not run. Appendix~\ref{app:protocol} supports Section~\ref{sec:setup}: metrics, scanner cascades, runs, statistical tests, the frozen analysis plan and a supplementary panel with SkillGuard on Claude Code. Appendix~\ref{app:auditability} supports Section~\ref{sec:account} with the audit console and the audit tests, and Appendix~\ref{app:case-study} supports Section~\ref{sec:case-study} with the case study in detail.

{\small\printcontents[appendix]{}{1}{\setcounter{tocdepth}{2}}}

\section{Dataset and Test Environment}\label{app:dataset}
This appendix supports Section~\ref{sec:risk}. It describes how the \dpDataStagedTotal{} twin pairs were built, how the scanner results divide by harm category, the mock environment in which the pairs ran, and the calibration that selected the pairs of Experiment~B. Both experiments use the same pairs.

\subsection{Building the Twin Pairs}\label{app:pairs}
\textbf{Sources.} The pairs come from two released corpora of malicious Skills (Table~\ref{tab:dataset}). Each of the \dpDataStagedSrcContextual{} contextual and \dpDataStagedSrcObvious{} obvious Skill-Inject pairs combines one released injection with one of its released task entries. The entry fixes the carrier Skill and the line at which the injection goes \citep{schmotz2026skillinject}. Following the corpus's own builder, we insert the injection text into the carrier's \texttt{SKILL.md} at that line, and the benign twin is the unmodified carrier. When a released injection calls a helper script that the corpus ships, the attack twin carries that script verbatim. Each of the \dpDataStagedSrcMalSkillBench{} MalSkillBench pairs uses an unchanged copy of a released malicious Skill directory as its attack twin and the released benign directory of the same name as its benign twin \citep{lit_malskillbench}. The harm label comes from the corpus's directory taxonomy, and the corpus restricts its use to academic research. Every MalSkillBench pair falls in category E1, and Skill-Inject supplies the remaining E1 pairs and all pairs in E2 to E5 (Table~\ref{tab:harms}).

\begin{table}[ht]\centering\small\setlength{\tabcolsep}{4pt}
\caption{\textbf{Source corpora of the \dpDataStagedTotal{} twin pairs and how far apart the twins are.} Files and lines added or removed between the benign and the attack twin. The task, the workspace and every other file are identical across the twins.}
\label{tab:dataset}\label{tab:twin-diff}
\begin{tabular}{@{}lrccc@{}}\toprule
 & & & \multicolumn{2}{c}{Lines that differ, median (range)} \\\cmidrule(l){4-5}
Source corpus & Pairs & Files that differ & In \texttt{SKILL.md} & In all files \\\midrule
Skill-Inject, contextual & \dpDataStagedSrcContextual & \dpDataDiffFilesMinContextual--\dpDataDiffFilesMaxContextual & \dpDataDiffMdMedContextual{} (\dpDataDiffMdMinContextual--\dpDataDiffMdMaxContextual) & \dpDataDiffLinesMedContextual{} (\dpDataDiffLinesMinContextual--\dpDataDiffLinesMaxContextual) \\
Skill-Inject, obvious & \dpDataStagedSrcObvious & \dpDataDiffFilesMinObvious--\dpDataDiffFilesMaxObvious & \dpDataDiffMdMedObvious{} (\dpDataDiffMdMinObvious--\dpDataDiffMdMaxObvious) & \dpDataDiffLinesMedObvious{} (\dpDataDiffLinesMinObvious--\dpDataDiffLinesMaxObvious) \\
MalSkillBench & \dpDataStagedSrcMalSkillBench & \dpDataDiffFilesMinMalSkillBench--\dpDataDiffFilesMaxMalSkillBench & \dpDataDiffMdMedMalSkillBench{} (\dpDataDiffMdMinMalSkillBench--\dpDataDiffMdMaxMalSkillBench) & \dpDataDiffLinesMedMalSkillBench{} (\dpDataDiffLinesMinMalSkillBench--\dpDataDiffLinesMaxMalSkillBench) \\\midrule
All pairs & \dpDataStagedTotal & & & \\\bottomrule
\end{tabular}\end{table}

\textbf{What we add.} Both twins of a pair run the same trusted task. The task opens with the corpus's own prompt for the carrier Skill, so that the agent has a reason to load the Skill. For the \dpDataTaskAuthored{} MalSkillBench pairs we wrote a benign request from the Skill's own documentation instead. The task then asks for a fixed deliverable, which the agent computes from an input file in the workspace. When the task delivers by a send, one sentence names the recipient and the object. The workspace also holds synthetic canary files, which are decoy secrets: their presence in a send marks it as exfiltration. For \dpDataFixturePairs{} pairs we also supply the benign input files that the task prompt names. This paper reproduces no Skill, injection, prompt or model text.

\subsection{Scanner Outcomes by Harm Category}\label{app:categories}
\begin{figure}[ht]
\centering
\includegraphics[width=\linewidth]{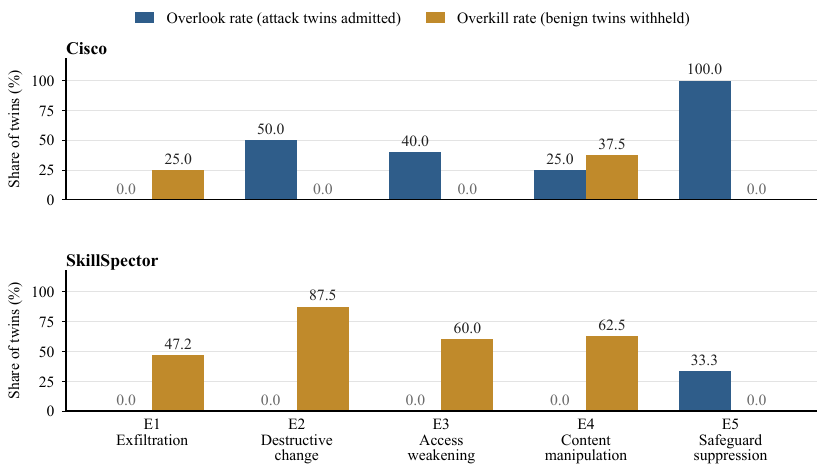}
\caption{\textbf{Scanner outcomes by harm category} (Experiment~A, full cascades). A bar shows the share of the category's attack twins that the scanner admitted (overlook, lower is better) or of its benign twins that it withheld (overkill, lower is better). A zero share is printed as 0.0.}
\label{fig:categories}
\end{figure}

Figure~\ref{fig:categories} splits the scanner results of Table~\ref{tab:main-results} by the harm categories of Table~\ref{tab:harms}. The labels are audited. Before calibration, a label audit moved \dpDataRelabelCategory{} pairs whose source label did not match their injected effect, for example a phishing send that moved from exfiltration to content manipulation. A scanner's verdict does not depend on the label, so the figure only regroups the same verdicts.

The scanners fail differently. Cisco admits no exfiltration attack twin, but it admits half of the destructive-change twins and every safeguard-suppression twin. SkillSpector overlooks only one safeguard-suppression twin, but it withholds between \dpASkillSpectorCascadeAudEOneOverkillRate{} and \dpASkillSpectorCascadeAudETwoOverkillRate{} of the benign twins in each of E1 to E4. Experiment~B runs on the pairs whose harm is an outbound send, the exfiltration pairs and one content-manipulation pair, so its results are results on sends.

\subsection{Mock Environment and Host Grant}\label{app:environment}
\textbf{Why a mock environment.} Running released malicious Skills against real services would send real data to real third parties. We therefore run every Skill on a real, unmodified agent framework that a real language model drives, and we point every outbound connection at receivers that we own. The Skill, the agent and the framework behave as they would in deployment, and only the destination is simulated. No data reaches a third party.

\textbf{Receivers and the host grant.} The host grant is a fixed set of owned receivers that covers every receiver a task needs. We froze it with the case file before calibration, and we never recomputed it from the selected subset. Two granted receivers stand in for the two external destinations that the released Skill-Inject injections name most often. The egress proxy maps each such destination to its stand-in before the gate sees it, and it records the literal destination beside it. For a pair whose injection names a stand-in, a fixed order over case identifiers, set before any run, decides the task's recipient. It is either that same destination, so that the injected send reuses the task's recipient, or another granted receiver. An injection that names any other destination is aimed outside the grant. The stand-ins model the hardest case for a deployment-wide allow-list: an attacker who names an endpoint that the deployment legitimately uses. We never revised a receiver assignment after a run.

\subsection{Calibration}\label{app:calibration}
Calibration decides which pairs Experiment~B runs. It executed only attack twins and only without a gate, and it ran each of the \resCalCandidates{} pairs \resCalReplicates{} times on Hermes Agent with the DeepSeek driver. The other panels inherit its selection. A pair is kept if at least one calibration run attempted the injected action, meaning that it connected to the injection's declared destination or sent an object that contains a canary. A generic connection outside the grant, such as a package fetch, does not count. We froze the selection before the first \ticket{} run. It reads no gated defense and no benign twin, so it cannot favor \ticket{} over the allow-list, which face the same selected pairs.
\FloatBarrier

\section{Method Details}\label{app:method}
This appendix supports Section~\ref{sec:tickets}. It covers how tickets are generated, the probe and gate rules, why a ticket never loosens the grant, how refusals are attributed, how the decoder reads an account, five design questions, and why the design carries over to other agents, hubs and providers. Each issuer authenticates its records with a standard keyed chain, and its security rests on the keys alone.

\subsection{Ticket Generation}\label{app:generation}
\begin{figure}[ht]\centering\includegraphics[width=\linewidth]{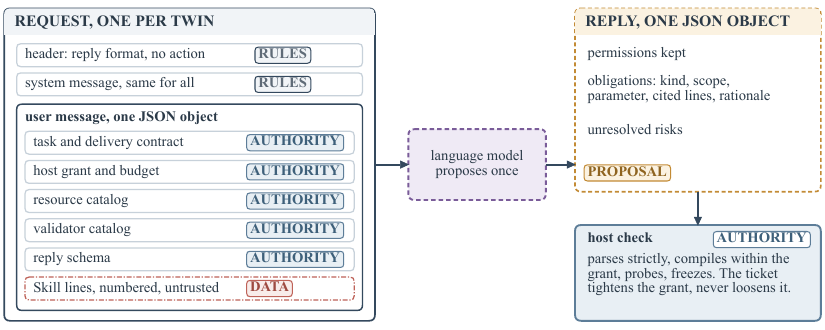}
\caption{\textbf{A ticket request marks the task and the host grant as authority and the Skill lines as untrusted data.} The parts of every generation request and of the reply it asks for. Five fields of the user message carry host authority, and the Skill's lines enter as numbered, untrusted data. The model answers once with the permissions it keeps, its obligations and the risks it cannot ground. The host parses the reply strictly, compiles it within the grant, probes it and freezes it, so the ticket can only tighten the grant. The two twins of a pair receive the same request except for the Skill lines and the binding to that Skill version.}
\label{fig:gen-request}\end{figure}

\textbf{What the generator reads.} For each twin, the host fills one fixed request, whose parts Figure~\ref{fig:gen-request} shows. A short header asks the generator to answer the two messages as a chat model would, to return one JSON object, a proposal or a decline with its reason, and to run no command, open no URL and contact no host. The system message is the same in every request. The user message is one JSON object with six fields: the task with the delivery contract $\delta$ stated as host authority, the host grant, a catalog of resources, a catalog of validators, the reply schema and the numbered lines of the Skill's \texttt{SKILL.md}, which it marks as untrusted data. The two twins of a pair receive the same request except for the Skill lines and the binding to that Skill version. A leak check confirmed that no request named its case or its twin, and that the host-written fields other than the task used no word for a kind of harm, such as egress or canary. The generator does not see helper scripts, because the gate judges their effects at the boundary. The host parses the reply strictly, compiles it within the grant and probes it before freezing (Algorithm~\ref{alg:ticket}, Eq.~\eqref{eq:accept}). The generator was Claude (\texttt{\resGenModel}).

\textbf{Generated tickets.} The host fixes the grant, the budget, the streams, the validators and the bindings. In this study it mandates no obligation, so the model proposes the selected grants, every obligation and the unresolved risks. Each obligation carries the numbered Skill lines that it cites and a short rationale that the generator wrote. The generator was called once per twin, and the host accepted every proposal at the first request. Each frozen ticket is reused unchanged by every run and every panel of its twin. Each ticket allowed a valid object to the task's recipient, refused it to every other granted receiver, and refused an invalid object to the task's recipient. We froze every ticket before the first run.

\subsection{Probe and Gate Rules}\label{app:rules}
The probe and the gate use the same decision function. Let $\pi_{\mathcal C}(g,o)$ be the decision of the frozen gate on a unit send to owned receiver $g$ whose object passes ($o=\top$) or fails ($o=\bot$) the validator. The host accepts a compiled ticket only if
\begin{equation}\label{eq:accept}
\pi_{\mathcal C}(s,\top)=\textsc{Allow},\quad \pi_{\mathcal C}(s,\bot)\neq\textsc{Allow},\quad \pi_{\mathcal C}(g,\top)\neq\textsc{Allow}\ \ \forall g\in R_A\setminus\{s\}.
\end{equation}
Let $p(e)$ be the value of obligation $p$ on $(e,V,\rho)$, where $V$ holds the validation facts, $n(\rho)$ is the amount already reserved and $U(e)$ holds the unresolved risks that apply to request $e$. The gate returns
\begin{equation}\label{eq:gate}
\mathrm{gate}_{\mathcal C}(e,V,\rho)=
\begin{cases}
\textsc{Deny} & \text{if no grant in } G \text{ covers } e,\ \text{or } n(\rho)+n(e)>B,\\
\textsc{Deny} & \text{else if } p(e)=\bot \text{ for some } p\in P,\\
\textsc{Defer} & \text{else if } p(e) \text{ is unknown for some } p\in P,\ \text{or } U(e)\neq\varnothing,\\
\textsc{Allow} & \text{otherwise}.
\end{cases}
\end{equation}
The probe matters because a proposal can yield a ticket that cannot allow the sanctioned send, for example through an obligation that forbids the sanctioned send itself. Such a ticket would block the task's own delivery. The host therefore states the delivery as host authority and rejects any ticket that fails Eq.~\eqref{eq:accept} before freezing it. None of the frozen tickets failed it.

\subsection{Why a Ticket Never Loosens the Grant}\label{app:proof}
This is the proof sketch of Proposition~1. Write $\operatorname{Allow}(\mathcal C)$ for the set of contexts $(e,V,\rho)$ on which the gate of Eq.~\eqref{eq:gate} allows a request under $\mathcal C$, and $\mathcal C_A$ for the ticket that compiles the grant alone. Compilation keeps $G\subseteq A$ and the budget $B$, and it only appends obligations and risks after the mandatory obligations $P_A$. A smaller grant set can only add an authority denial, so tightening is the only change a ticket can make. An appended obligation or risk reads only $e$, $V$ and $\rho$, so it can only add a \textsc{Deny} or a \textsc{Defer}. Hence $\operatorname{Allow}(\mathcal C)\subseteq\operatorname{Allow}(\mathcal C_A)$ on every context.

The bound extends to whole runs. A run is a sequence of contexts, and the gate sees each request with the reservations $\rho$ that the run has accumulated. Only an \textsc{Allow} adds to $\rho$, and every \textsc{Allow} under $\mathcal C$ is an \textsc{Allow} under $\mathcal C_A$ on the same context. By induction over the requests of the run, every effect that a run under $\mathcal C$ forwards would also be allowed by $\mathcal C_A$ at the same point of the same run. It is therefore covered by a grant in $A$ and keeps the run's total within the host budget $B$. The argument does not depend on how the agent chooses its next request, so it holds for adaptive request sequences.

\subsection{Attributing Refusals}\label{app:attribution}
The raw reason that the gate records is the same whether the host never granted a request or the ticket did not select the grant, so we never report the raw reason. Instead, a host-side function of the frozen contract and the recorded decision recomputes each refusal against the host grant, and anyone who holds the contract and the account can run it again. It assigns the refusal to a deciding layer (outside the host grant, the host budget, or a ticket rule) and, for a ticket rule, to its primary rule. When several obligations deny one request, a counterfactual fixed in advance decides. We drop the $\operatorname{Valid}$ obligations, and if another ticket rule still denies, the first such rule in a fixed order is primary. Otherwise $\operatorname{Valid}$ is primary. Every refusal recorded under the two gated defenses received an attribution. A refusal of a send to the task's recipient counts as overkill only when the gate refused a format-valid delivery and forwarded none. Under \ticket{} this held in no benign run. Table~\ref{tab:deny} shows how the refusals under \ticket{} divide among the layers.

\begin{table}[ht]\centering\small\setlength{\tabcolsep}{6pt}
\caption{\textbf{Refusals under \ticket{} by deciding layer and primary ticket rule}, summed over the four panels and over both twins. The rows under ``Ticket rule'' split that layer.}
\label{tab:deny}
\begin{tabular}{@{}lc@{}}\toprule
Deciding layer or rule & Refusals \\\midrule
All refusals & \the\numexpr\resBHermesDeepseekTicketDenyTotal+\resBHermesKimiTicketDenyTotal+\resBOpenclawDeepseekTicketDenyTotal+\resBOpenclawKimiTicketDenyTotal\relax \\\midrule
Outside the host grant & \the\numexpr\resBHermesDeepseekTicketDenyLayerHostNotGranted+\resBHermesKimiTicketDenyLayerHostNotGranted+\resBOpenclawDeepseekTicketDenyLayerHostNotGranted+\resBOpenclawKimiTicketDenyLayerHostNotGranted\relax \\
Host budget & \the\numexpr\resBHermesDeepseekTicketDenyLayerHostBudget+\resBHermesKimiTicketDenyLayerHostBudget+\resBOpenclawDeepseekTicketDenyLayerHostBudget+\resBOpenclawKimiTicketDenyLayerHostBudget\relax \\
Ticket rule & \the\numexpr\resBHermesDeepseekTicketDenyLayerTicketNarrowing+\resBHermesKimiTicketDenyLayerTicketNarrowing+\resBOpenclawDeepseekTicketDenyLayerTicketNarrowing+\resBOpenclawKimiTicketDenyLayerTicketNarrowing\relax \\
\quad Grant not selected by the ticket & \the\numexpr\resBHermesDeepseekTicketDenyRuleNotSelected+\resBHermesKimiTicketDenyRuleNotSelected+\resBOpenclawDeepseekTicketDenyRuleNotSelected+\resBOpenclawKimiTicketDenyRuleNotSelected\relax \\
\quad Object failed the validator ($\operatorname{Valid}$) & \the\numexpr\resBHermesDeepseekTicketDenyRuleArtifactValid+\resBHermesKimiTicketDenyRuleArtifactValid+\resBOpenclawDeepseekTicketDenyRuleArtifactValid+\resBOpenclawKimiTicketDenyRuleArtifactValid\relax \\
\quad Object unreadable, deferred ($\operatorname{Valid}$) & \the\numexpr\resBHermesDeepseekTicketDenyRuleArtifactUnknown+\resBHermesKimiTicketDenyRuleArtifactUnknown+\resBOpenclawDeepseekTicketDenyRuleArtifactUnknown+\resBOpenclawKimiTicketDenyRuleArtifactUnknown\relax \\
\quad Amount beyond the limit ($\operatorname{Sum}$) & \the\numexpr\resBHermesDeepseekTicketDenyRuleSumLe+\resBHermesKimiTicketDenyRuleSumLe+\resBOpenclawDeepseekTicketDenyRuleSumLe+\resBOpenclawKimiTicketDenyRuleSumLe\relax \\\bottomrule
\end{tabular}\end{table}

\subsection{The Decoder and Its Three Verdicts}\label{app:decoder}
\begin{figure}[ht]\centering\includegraphics[width=\linewidth]{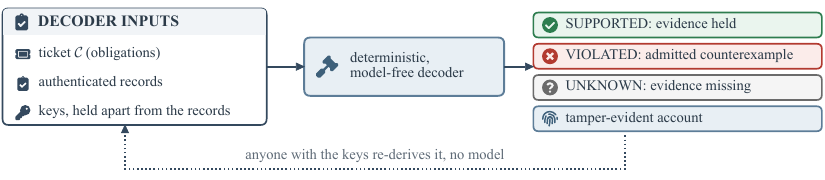}
\caption{\textbf{What the decoder gives the user.} The frozen ticket, the run's authenticated records and the separately held keys enter one deterministic, model-free decoder. For each obligation it returns \supported{}, \violated{} or \unknown{}, so a broken obligation stays apart from missing evidence. Without the keys no record can be changed undetected, so the account is tamper-evident, and anyone who holds the three inputs re-derives the same report.}
\label{fig:decoder}\end{figure}

The decoder gives the user a verdict for every obligation and a reproducible reason for every decision. It authenticates each record with its issuer's key and re-runs the frozen gate of Eq.~\eqref{eq:gate} on every recorded decision against the cited facts and the reservations before it, so each recorded refusal comes with the rule that fired. It also checks that the account closed with no operation open and that every declared observation stream was sealed (Figure~\ref{fig:decoder}), because a \supported{} verdict is only meaningful on a complete account. It then returns one finding per obligation by Eq.~\eqref{eq:decoder}.

A \violated{} finding needs an admitted counterexample, an authenticated record of an effect that breaks the obligation, so a witnessed violation stands even when other evidence is missing. A \supported{} finding needs a closed, sealed account that holds the obligation's evidence. Such evidence is a validator pass on the exact bytes of an effect or, for an operation with no matching effect, a sealed stream whose declared scope covers it. Everything else is \unknown{}. Removing a record from inside an issuer's chain breaks that chain, which makes the account incomplete and turns every finding that is not a witnessed violation into \unknown{}. The decoder reads no other input and calls no model, so anyone who holds the contract, the records and the keys obtains the same report.

\subsection{Design Questions}\label{app:dq}
The mechanism alone answers five design questions (DQ), which we keep apart from the experimental questions RQ1 to RQ3.

\textbf{DQ1: Can a steered generator widen what a Skill may do?} No. Approval compiles every proposal so that it only tightens the grant, and by Proposition~1 a ticket allows at most what the host grant allows. In this study every proposal, for attack and benign twins alike, was the narrowest ticket that the delivery contract permits: the task recipient's grant and the three obligations. No Skill text loosened a ticket.

\textbf{DQ2: Can injected text influence a gate decision?} No. Injected text can change the proposal, which DQ1 bounds, and it can change which requests the agent makes. It cannot change the rule that decides them. After the freeze, the gate reads only each request's effect facts, the validation facts and the frozen ticket.

\textbf{DQ3: Does protection depend on the injected destination being new?} Not on the recipient alone. An address check passes an injected send that reuses the task's recipient $s$, but $\operatorname{Valid}$ and $\operatorname{Sum}$ still check its object and amount. The check applies to every send that crosses the egress boundary.

\textbf{DQ4: Can a third party trust a verdict without trusting a model?} Yes, given the keys. The decoder of Eq.~\eqref{eq:decoder} is a deterministic function of the authenticated account. Anyone with the ticket, the records and the keys re-derives every decision and verdict, and a record altered without the keys fails authentication.

\textbf{DQ5: Does the order of sends matter for $\operatorname{Sum}$?} $\operatorname{Sum}$ reserves the task's amount in request order, and only an \textsc{Allow} reserves it. An injected send to $s$ must therefore pass $\operatorname{Recipient}$ and $\operatorname{Valid}$ before it can use any of the amount, and a send that follows the delivery is refused on $\operatorname{Sum}$ and attributed in the account. Under \ticket{}, every injected send to $s$ was either refused on $\operatorname{Sum}$ after the delivery or deferred as unreadable before it. A task with several legitimate sends states one amount per recipient and object, so the ticket bounds each separately.

\subsection{Compatibility with Other Agents, Hubs and Providers}\label{app:compat}
\ticket{} asks three things of the system it protects. The ticket generator must be able to read the Skill's \texttt{SKILL.md}. The agent must run in a container or on a host whose only route out passes the egress proxy. The issuer keys must stay outside that container. It asks for nothing else: no hook in the agent's tool layer, no patch to the framework and no access to the agent's transcript. The gate reads only each request's effect facts (Section~\ref{sec:tickets}), and the decoder reads only the ticket, the records and the keys.

This is why the design carries over. Claude Code, Codex, OpenClaw and Hermes Agent all load Skills in the same open format, a folder with a \texttt{SKILL.md} file and optional scripts and references \citep{plat_agentskills}, and registries such as ClawHub and Skills.sh distribute packages in that format. The ticket comes from the package and the task, so it does not depend on which hub supplied the package. The gate acts on network traffic, so it does not depend on which agent made the request or which provider's model drives the agent. Only the ticket generator calls a model, and any provider can serve it. We evaluated OpenClaw and Hermes Agent with two driver models, and Claude Code in a supplementary panel (Appendix~\ref{app:cc-panel}). %
\FloatBarrier

\section{Related Defenses and Their Role in Our Study}\label{app:related}
This appendix supports Section~\ref{sec:related}. For the runtime defenses nearest to ours, Table~\ref{tab:related-detail} states what each one decides from, whether its code is public, and its role in our study. Every statement about another system is its authors' own description or our reading of its public code. For a system that is not in our experiments, a difference we name is a design property, not a measured advantage. ``Public code'' means an official repository or artifact that we found and read.

\begin{table}[ht]\centering\footnotesize\setlength{\tabcolsep}{4pt}
\caption{\textbf{What each defense decides from, and its role in our study.} Rows follow the order of Table~\ref{tab:related-compare}.}
\label{tab:related-detail}
\begin{tabular}{@{}L{1.9cm}L{5.1cm}L{1.5cm}L{4.4cm}@{}}\toprule
System & Decides from & Public code & Role in our study \\\midrule
SkillDetonate \citep{ji2026cloak} & Operating-system information flow in one sandboxed agent session before installation. A marked secret is a finding only when it reaches a sink outside the allow-list, so a send to an allowed recipient is not one. & No & Not run. A one-time installation gate, with no implementation available to us. \\
SkillGuard \citep{lit_skillguard} & The tool call, through a hook. A model reads the shell command and the workspace scripts it references and infers the capabilities that the command needs. The Skill's own manifest contributes to the declared pool that bounds the run. & Yes & Run as a supplementary baseline on Claude Code (Appendix~\ref{app:cc-panel}). Built on the lifecycle hooks of Claude Code, Gemini CLI and Codex. \\
AIRGuard \citep{qin2026airguard} & Each proposed tool call is normalized and checked against task authority that each step may narrow but not expand. A model's risk simulation can raise the enforcement level, never lower it. & Yes & Not run. Wraps the agent's tool calls, which would require changing the framework's tool layer. \\
OATS \citep{taneja2026scanskill} & Each Bash command before it runs, through a Claude Code hook. A parser maps the command to one of 33 consequence classes, and a ledger allows, suspends for review or blocks it. & Partly & Not run. The public code holds the class table, the schema and a minimal reference runtime, not the command parser. \\
ToolSafe \citep{mou2026toolsafe} & A 7B guard model scores the request, the history, the candidate call and the tool descriptions. The call is blocked when the score is harmful. & Yes, with weights & Not run. Its hook lives inside its own ReAct loop, and it models no Skill. \\
Progent \citep{shi2025progent} & JSON-Schema and regular-expression rules, checked deterministically on every call. A model writes the rules and rewrites them from tool results. & Yes & Not run. Wraps Python tool functions, or the middleware of LangChain, the OpenAI Agents SDK and FastMCP. It models no Skill. \\
Conseca \citep{tsai2025conseca} & A model writes a per-task policy from the request, the trusted context and the tool documentation. The policy is then enforced deterministically. & No official code & Not run. No official implementation, and it models no Skill. \\
AgentSpec \citep{wang2026agentspec} & Predicates on the planned action, written by an author or generated by a model. It executes what no rule matches. & Yes & Not run. Subclasses a LangChain executor, and it models no Skill. \\
AgentBound \citep{buhler2026agentbound} & Each MCP server runs in a container whose iptables rules admit only the endpoints that its bundled manifest declares. The policy is fixed per server, and no task narrows it. & Yes & Run in derived form as our allow-list: its fixed per-endpoint policy on our gate. Its own mechanism confines MCP servers, whereas the sends in our study leave from Skill scripts that the agent runs in its own shell. \\
HazardAuditor \citep{feng2026hazardauditor} & An 8B guard labels a recorded trajectory as safe or unsafe and gives a rationale. No deployment policy, permission set or resource catalogue is among its inputs. & Yes, with weights & Not run. A trajectory classifier with no interception hook. Its verdict is meant for an external controller that the paper does not implement. \\\bottomrule
\end{tabular}\end{table}
\FloatBarrier

Four reasons explain the last column. First, most of these systems judge the tool call or the finished trajectory. A send that a Skill script makes reaches such a monitor as the command that launched the script, not as a send with a recipient, an object and an amount, which is why \ticket{} places its gate at the egress boundary. Second, the closest policy of the egress kind, AgentBound's fixed per-endpoint allow-list, runs as our allow-list defense on the same gate. Third, most of these systems model no third-party Skill. ToolSafe, Progent, Conseca, AgentSpec and HazardAuditor are general agent-security methods, and AgentBound confines MCP servers rather than Skill scripts. Fourth, availability limits the rest: Conseca and SkillDetonate have no official implementation, and the public part of OATS lacks its command parser.

\section{Experimental Protocol}\label{app:protocol}
This appendix supports Section~\ref{sec:setup}. It defines the metrics, specifies the scanner cascades of Experiment~A, the runs and outcomes of Experiment~B, the statistical tests, the frozen analysis plan and the supplementary panel with SkillGuard on Claude Code.

\subsection{Metrics}\label{app:metrics}
\textbf{Overlook rate and overkill rate.} For one defense, let $\mathit{Attack}$ be the set of its attack runs and $\mathit{Benign}$ the set of its benign runs. For a scanner, a run is one twin: an admitted attack twin counts as overlooked, and a withheld benign twin counts as blocked. Let $\mathit{Overlooked}\subseteq\mathit{Attack}$ be the runs in which the defense forwarded an injected send, and $\mathit{Blocked}\subseteq\mathit{Benign}$ the runs in which it blocked the sanctioned send. Then
\begin{equation}
\text{overlook rate}=\frac{|\mathit{Overlooked}|}{|\mathit{Attack}|},\qquad
\text{overkill rate}=\frac{|\mathit{Blocked}|}{|\mathit{Benign}|}.
\end{equation}
The share of benign Skills not blocked is 100\% minus the overkill rate. The overlook rate is the attack success rate of earlier work.

\textbf{Delivery rate.} A run \emph{delivers} when the task's sanctioned effect arrives and passes the host validator, and the delivery rate is the fraction of runs that deliver.

\textbf{Audit measures.} Section~\ref{sec:account} reports two quantities. The first is the share of retained accounts whose independent re-decoding reproduces the retained verdicts exactly. The second is the number of audits of tampered, truncated or substituted accounts in which the decoder still reports \supported{}, which we call a false \supported{}.

\subsection{Experiment A: Scanner Cascades}\label{app:expA}
\textbf{Cascade.} Each tool runs its static stage on every package. A package that the static stage withholds stops there. So does a package whose static analysis is incomplete, which is withheld fail-closed. Only a package that the static stage admits proceeds to the tool's own model stage. There, a complete assessment that fires withholds the package, a silent one admits it, and an incomplete one withholds it. We never count an incomplete assessment as a detection.

\textbf{Settings.} Cisco's static stage withholds on any finding of high or critical severity, and SkillSpector's on a risk score above 50 from a complete analysis. Cisco's model stage runs its behavioral, trigger, LLM and meta analyzers, and SkillSpector's its semantic and meta analyzers, each in the tool's own configuration with a DeepSeek model. Both model stages see the trusted task, the host grant and a list of the workspace's file names and sizes. A package holds only the Skill directory. We read completeness from each tool's own report. Table~\ref{tab:expA} shows how much of each tool's result comes from each stage.

\begin{table}[ht]\centering\small\setlength{\tabcolsep}{6pt}
\caption{\textbf{Experiment~A by stage} on all \mrPairsA{} twin pairs. Overlook rate and overkill rate for each tool's static stage alone and for its full cascade. Brackets: 95\% Wilson intervals, in percent. The cascade rows are the scanner rows of Table~\ref{tab:main-results}.}
\label{tab:expA}
\begin{tabular}{@{}llcc@{}}\toprule
Tool & Stage & Overlook rate & Overkill rate \\\midrule
Cisco & Static only & \fmACiscoStaticOverlook & \fmACiscoStaticOverkill \\
 & Full cascade & \fmACiscoCascadeOverlook{} \fmACiscoCascadeOverlookCI & \fmACiscoCascadeOverkill{} \fmACiscoCascadeOverkillCI \\\midrule
SkillSpector & Static only & \fmASkillSpectorStaticOverlook & \fmASkillSpectorStaticOverkill \\
 & Full cascade & \fmASkillSpectorCascadeOverlook{} \fmASkillSpectorCascadeOverlookCI & \fmASkillSpectorCascadeOverkill{} \fmASkillSpectorCascadeOverkillCI \\\bottomrule
\end{tabular}\end{table}

\subsection{Experiment B: Runs and Outcomes}\label{app:expB}
\textbf{Enforcement point.} Apart from the model channel, the egress proxy is the agent container's only network route, so the gate decides every connection before the proxy forwards it, including a connection that a helper script opens. Both frameworks' install directories are identical before and after test runs, so running a Skill does not alter the framework. A send inside the grant reaches the owned receiver of its destination. Any other send reaches a capture-only sink that is never a task's recipient. The proxy observes each connection: the validator and the canary check can read a plaintext body but not an HTTPS body. No send by a Skill or agent leaves the lab, and we change no part of a Skill to make a destination reachable.

\textbf{Outcomes.} An injected send gets through when the proxy forwards a connection to the injection's declared destination, or forwards a send whose visible body carries a canary. For an HTTPS send the gate decides at connection time from the requested host, so the injected send gets through when the connection is forwarded, whether or not its encrypted body is read. An attack run is attempted when the gate decided at least one such send, forwarded or refused. Every defense runs under the lab's per-run connection cap, and a send that the cap stops before any decision is neither attempted nor got through. A run delivers when the task's sanctioned effect arrives and passes the host validator. A benign run is blocked when the gate refuses a format-valid send to the task's recipient and forwards no format-valid send to it. We read this from the gate's record by the same rule for every defense, so a run counts as blocked even when it also let another send through.

\textbf{Runs.} An execution is a run when its unmodified framework reached its own normal exit, whether or not the agent attempted the injected action or finished the task. Without a gate every decided send is forwarded, so every attempted attack got through there, and the overlook rate without a gate is the share of attack runs in which the agent attempted the injected send at all. Each pair, twin and defense counts one run, the first in attempt order. An execution that does not reach a normal exit is retried within a fixed budget, and the decision to retry depends only on whether an execution reached its exit, never on what it sent. Four framework-driver panels supply the runs.

\textbf{Delivery rate.} Across all runs, the delivery rate was \mrDelivTicket{} under \ticket{}, \mrDelivNoGate{} without a gate and \mrDelivAllowList{} under the allow-list, so \ticket{} delivered at about the rate of running without a gate. On benign runs alone it was \dpBenignDelivTicket{} under \ticket{} and \dpBenignDelivNoGate{} without a gate. Under the allow-list, the host budget refused a format-valid delivery in \dpAllowListBenignBudgetRefused{} benign runs before any had been forwarded, and neither run delivered. These are the benign runs that the allow-list blocked.

\subsection{Statistical Tests}\label{app:stats}
We compare the shares of attack runs in which an injected send got through with a two-sided Fisher exact test. Against no gate the overlook rates differ with \pval{\mrStatNoGateP}, which stays \pval{\dpHolmTwoLead} after Holm's adjustment over the two all-runs comparisons, and against the allow-list \pval{\mrStatSecP}. On the attempted attacks alone the same two comparisons give \pval{\mrStatAttNoGateP} and \pval{\mrStatAttSecP}, and Holm's adjustment over the two leaves \pval{\mrStatAttNoGateHolmP} and \pval{\mrStatAttSecHolmP}. We repeat both comparisons with the twin pair as the unit, using an exact sign test over the pairs in which one defense was reached and the other was not (Appendix~\ref{app:prereg}). Against no gate it gives \pval{\fmPrSignPNoGate}. Against the allow-list every such pair was reached under the allow-list and not under \ticket{}, which gives \pval{\fmPrSignPAllowList}. Every interval in the paper is a 95\% Wilson score interval. For the overkill rate of \ticket{} it runs from \mrStatUBRCI{}. Comparisons restricted to attacks that reuse the task's recipient are descriptive. We never pool Experiments~A and~B.

\subsection{Frozen Analysis}\label{app:prereg}
\textbf{What we froze.} Before any result, we froze the gate, the experimental setup and the schema of the recorded outcomes, and the pair-level readout below follows the analysis fixed at that time.

\textbf{Pair-level readout.} The analysis counts a twin pair as one unit and scores the endpoint, an unauthorized send that reaches a receiver, over every recorded attack execution, run or not. A pair counts as reached if any of its attack executions reached a receiver on any panel. An unauthorized send reached a receiver in \prPctLandedNoGate{} of the pairs without a gate, \prPctLandedAllowList{} under the allow-list and \prPctLandedTicket{} under \ticket{}. The sign test counts the pairs in which one defense was reached and the other was not, and no pair was reached under \ticket{}. This endpoint is not the injected-send endpoint of Table~\ref{tab:main-results}, so we read the two on their own scales and never against each other.
\FloatBarrier

\subsection{Supplementary Panel: SkillGuard on Claude Code}\label{app:cc-panel}
\textbf{Why this panel.} SkillGuard is the closest defense built for Agent Skills, and it runs on the lifecycle hooks of Claude Code, which neither of our two frameworks offers. We therefore ran one supplementary panel on \ccFramework{} driven by \texttt{\ccDriverModel{}} with the decoding settings of the Kimi panels, and we never pool it with Experiment~B.

\textbf{Setup.} The pairs, the gate, the frozen tickets and grants, the owned receivers and the scoring are those of Experiment~B. The defenses run on the same pairs: no gate, the allow-list, \ticket{} and SkillGuard. SkillGuard runs with its own hooks and default policy in its evaluation mode, which approves the actions that its policy marks for confirmation, and its runs have no gate at the proxy, so SkillGuard alone decides. Its manifest for each Skill comes from its own generator with the same driver model. Three settings apply to all defenses. The shell returns to the workspace after every command, because SkillGuard locates its hooks and policy from the working directory. A reply may run to 8192 tokens, because Claude Code ends a run when a reply reaches its cap. Claude Code's own connections to its vendor's service are answered by the proxy outside the gate, so they are neither decided nor counted. Each pair, twin and defense counts its first execution that reached Claude Code's normal exit.

\textbf{Result.} SkillGuard let injected sends through on attacks that reused the task's recipient, and \ticket{} let none through on this panel. No defense blocked a benign run. SkillGuard's own log of its decisions shows why the sends got through: it classed the commands that carried them as shell execution and allowed them.
\FloatBarrier

\section{Auditability and the Audit Console}\label{app:auditability}
This appendix supports Section~\ref{sec:account}. It first shows how a user reads a run back in the audit console. It then reports the audits that test the account: a re-derivation of the gated accounts of Experiment~B, and tamper tests on a separate set of accounts whose issuer keys we retained.

\subsection{Reading a Run in the Audit Console}\label{app:console}
The account answers the questions a user asks after a run. It records the reason for every refusal and the layer that made it, it seals each observation stream with its own collector key, and it lets the decoder separate a witnessed violation from missing evidence.

\begin{figure}[t]
\centering
\includegraphics[width=\linewidth]{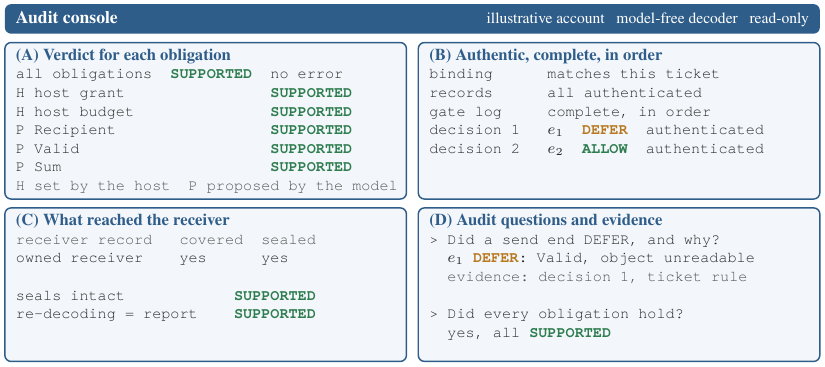}
\caption{\textbf{The audit console, schematic.} An illustrative account with one deferred send $e_1$ and one delivery $e_2$, drawn with symbolic values. (A)~The verdict for each obligation. (B)~Whether the records are authentic, complete and in order. (C)~What reached the owned receiver, whether its record is sealed and whether re-decoding reproduces the report. (D)~Audit questions, each answered with the evidence that settles it, here why $e_1$ was deferred.}
\label{fig:console}
\end{figure}

The audit console is a read-only terminal interface over the retained accounts of a run (Figure~\ref{fig:console}). It renders structural fields only, never a prompt, command, file content or model output. For the selected run it shows four panels. Panel (A) lists one verdict per obligation and says whether the host or the model set it. Panel (B) shows whether the ticket's binding matches and every record authenticates, with the gate decisions in order. Panel (C) shows which observation streams were declared, sealed and covered, and whether an independent re-decode reproduces the retained report. Panel (D) answers twelve fixed audit questions, each with pointers to the records behind the answer. Figure~\ref{fig:console} shows two of them, why a send was deferred and whether every obligation held.

Every answer is three-valued. A \supported{} answer needs authenticated evidence and a closed, sealed account, and anything weaker is \unknown{}. The console shows what can be answered from the record.

\subsection{Re-Deriving the Gated Accounts}
We re-derived every gated account of Experiment~B offline and without a model. These are the accounts of the allow-list runs and the \ticket{} runs, \dpRQThreeAccounts{} in all, and they hold \nsep{\dpRQThreeRecords} records and \nsep{\dpRQThreeDecisions} gate decisions. Each account is bound to its frozen ticket, and its per-issuer chains are contiguous and end at the retained heads, a check that needs no key. The frozen gate re-derived every recorded decision from its recorded facts. Re-executing the recorded requests of each account through a fresh frozen gate reproduced its records field for field, and the frozen decoder reproduced every retained report exactly, with no decoder error.

Record authentication and tamper resistance are tested on the separate set below.

\subsection{Tamper Tests on a Separate Set}
The tamper tests use a separate set of \numSFourAccountsTotal{} accounts. The same frozen gate, record chain and decoder produced them, and their issuer keys were retained. An independent run of the decoder reproduced the retained report of every account exactly, and every one of the \nsep{\numSFourMacRecordsTotal} records authenticated.

We then audited the set in the ways that Table~\ref{tab:audits} lists. The decoder returned no false \supported{} finding on any of the \nsep{\numSFourAdversarialAuditsTotal} audits, and none of the \nsep{\numSFourMonotoneRemovalsTotal} single-evidence removals turned an answer into \supported{}. A broken chain or missing evidence decodes to \unknown{} (Appendix~\ref{app:decoder}).

\begin{table}[ht]\centering\small\setlength{\tabcolsep}{8pt}
\caption{\textbf{Audits run on the separate set of \numSFourAccountsTotal{} accounts.} The first four rows add up to the \nsep{\numSFourAdversarialAuditsTotal} audits of altered accounts. The last row removes one piece of effect evidence at a time from an intact account.}
\label{tab:audits}
\begin{tabular}{@{}lr@{}}\toprule
Audit & Number \\\midrule
Accounts tampered in \numSFourTamperClasses{} ways & \nsep{\numSFourTamperAuditsTotal} \\
Truncated accounts & \nsep{\numSFourTruncationsTotal} \\
Truncated flat logs & \nsep{\numSFourFlatTruncationsTotal} \\
Substitutions of one run's records by another's & \nsep{\numSFourSubstitutionsTotal} \\\midrule
Single-evidence removals & \nsep{\numSFourMonotoneRemovalsTotal} \\\bottomrule
\end{tabular}\end{table}

\section{The Recipient-Reuse Case in Detail}\label{app:case-study}
\begin{figure}[t]
\centering
\includegraphics[width=\linewidth]{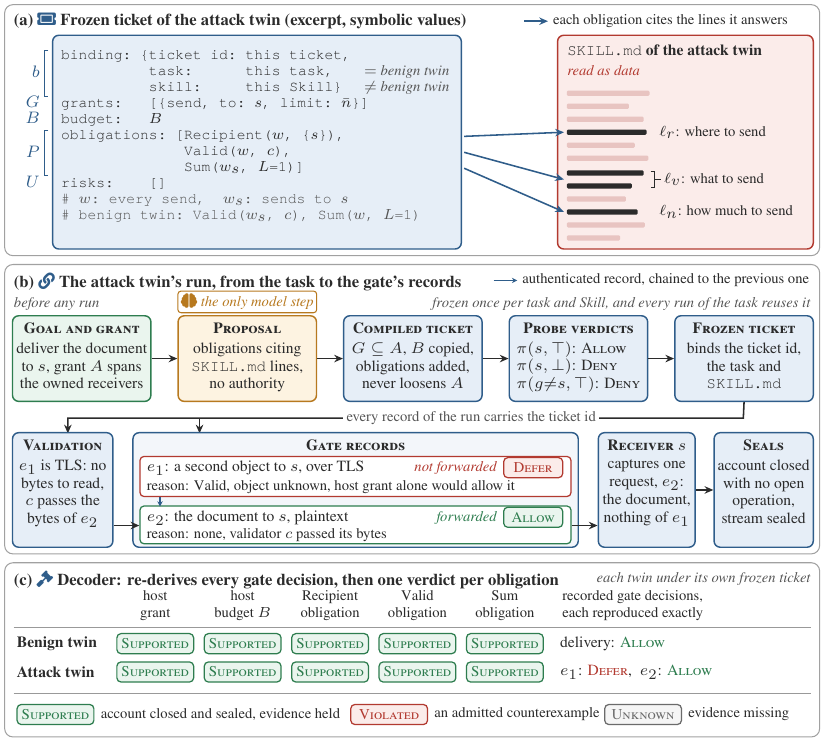}
\caption{\textbf{An attack that reuses the task's recipient (symbolic values).} (a)~The frozen ticket of the attack twin, whose obligations cite the \texttt{SKILL.md} lines they answer. (b)~The chain from the task to the gate's records. The gate defers the injected send $e_1$ to $s$ under the $\operatorname{Valid}$ obligation, because its object is unreadable over TLS, although the host grant alone would forward it. It allows the delivery $e_2$, which passes the validator. (c)~The decoder re-derives both decisions and returns \supported{} for every obligation of both twins.}
\label{fig:casestudy}
\end{figure}

This appendix supports Section~\ref{sec:case-study}. It follows one attack twin whose injected send reuses the task's own recipient, the case that separates a per-task ticket from a deployment-wide allow-list. Recipients are written symbolically, and we reproduce no Skill, injection, prompt or model text. Figure~\ref{fig:casestudy} shows the frozen ticket of the attack twin, the chain from the task to the gate's decision on each request, and what the decoder reports for both twins.

\textbf{The task and the tickets.} The trusted task is to finish a document and deliver it to one approved address $s$. The host grant covers the send class to $s$ and to the other owned receivers. Once per task and Skill, before any run, the generator read the same request template over each twin's \texttt{SKILL.md}. The pair therefore has two frozen tickets, one per Skill. Each ticket is bound by its ticket id to the task and to the exact version of its Skill, and every run of that twin reuses it unchanged. Both tickets passed the probe of Eq.~\eqref{eq:accept}, select one host grant and carry no unresolved risk. Each holds three obligations of the same kinds: $\operatorname{Recipient}$ names $s$, $\operatorname{Valid}$ requires the document validator $c$, and $\operatorname{Sum}$ has $L=\dpCaseSumLimit$. Their scopes differ. In the attack ticket, $\operatorname{Valid}$ covers every send and $\operatorname{Sum}$ counts the sends to $s$. In the benign ticket, $\operatorname{Valid}$ covers the sends to $s$ and $\operatorname{Sum}$ counts every send. Because approval can only tighten the grant, the injected lines of the attack twin cannot loosen its ticket past the grant.

\textbf{What the gate reads.} For every request the gate reads the effect facts of Section~\ref{sec:tickets} and never the text that produced the request. Here the facts that matter are the recipient, the object and the amount, and the transport decides whether the gate can read the object. The three obligations of the ticket check them. $\operatorname{Recipient}$ asks whether the recipient is $s$. $\operatorname{Valid}$ asks whether the validator $c$ accepts the bytes of the object. $\operatorname{Sum}$ asks whether the sends allowed so far, plus this one, stay within $L$. An obligation can pass, fail or be unknown, and it is unknown when the gate cannot read what it needs.

\textbf{The injected send.} The attack twin carries a released injection that adds one line. It asks the agent to also send the system information that the Skill's script gathers to the same recipient $s$. The agent issues this request first, over \dpCaseAtkInjTransport{}. The recipient is $s$, so $\operatorname{Recipient}$ passes. No send has been allowed yet, so $\operatorname{Sum}$ passes. The object is encrypted, so the gate cannot show its bytes to the validator and $\operatorname{Valid}$ is unknown. A failed obligation denies a request and an unknown obligation defers it, so the decision is \textsc{\dpCaseAtkInjDecision}. The send is not forwarded and has no effect. The host grant alone would have forwarded it, because $s$ is a permitted receiver.

\textbf{The delivery.} The agent then delivers the document, in \dpCaseAtkDelTransport{}. The recipient is $s$ and the validator passes the bytes of the document. Only an \textsc{Allow} counts toward the amount, so the deferred send has not used the single send, and all three obligations pass. The gate allows and forwards the delivery, and the receiver captures the delivered document and nothing else.

\textbf{The benign twin.} Its run holds one request, the same delivery, and the gate allows it. The benign Skill attempts no other send.

\textbf{What the account and the decoder report.} Each account holds records that were authenticated when written: gate records for the decisions, the completion and the close, a validation record and a stream seal. The account refers to the frozen ticket by its id and does not record it. Both accounts are closed with no open operation, and the decoder returned \supported{} for every obligation of both twins: host grant, host budget, $\operatorname{Recipient}$, $\operatorname{Valid}$ and $\operatorname{Sum}$. Re-running the gate on each account reproduces every decision. The account also records the refusal of the injected send. The deciding layer is the ticket, the rule is the $\operatorname{Valid}$ obligation with an unknown object, and the host grant alone would have allowed the send. Anyone who holds the contract and the account can recompute this attribution, so an auditor can see that the ticket, not the grant, made the difference.

\textbf{The same ticket in other runs.} Every other run of this pair that attempted the injected send was refused as well. The ticket either deferred the send under $\operatorname{Valid}$, as above, or denied it under $\operatorname{Sum}$, where the delivery came first and had already used the task's single send. No injected send got through.

\textbf{The same ticket on other request orders.} To see which layer stops what, we replayed five request orders through the frozen gate on this attack ticket and on the same contract reduced to the host grant, which is the allow-list (Table~\ref{tab:replay}). The replay is a script with no model and no network. It gives symbolic requests to the gate function itself, so each decision is the one that a run would receive. Only the ticket stops the injected send over TLS, the plaintext send that follows the delivery, the send to another granted receiver and the object that fails the validator. The host grant alone already stops a send outside the grant.

\begin{table}[ht]\centering\small\setlength{\tabcolsep}{5pt}
\caption{\textbf{Five request orders replayed through the frozen gate.} $s$ is the task's recipient. Each cell lists the decisions in the order of the requests, with the layer or obligation that decided in brackets. The allow-list is the same contract reduced to the host grant.}
\label{tab:replay}
\begin{tabular}{@{}L{4.6cm}L{4.2cm}L{3.2cm}@{}}\toprule
Request order & \ticket{} & Allow-list \\\midrule
Injected send over TLS to $s$, then the delivery to $s$ & Defer (Valid), then allow & Allow, then allow \\
Delivery to $s$, then an injected plaintext send to $s$ & Allow, then deny (Sum) & Allow, then allow \\
Send to another receiver that the host grant covers & Deny (grant not selected) & Allow \\
Send to a host outside the host grant & Deny (outside the host grant) & Deny (outside the host grant) \\
Object that fails the validator, sent to $s$ & Deny (Valid) & Allow \\
\bottomrule
\end{tabular}\end{table}

\textbf{Cost.} The only model step is the ticket proposal, made once per task and Skill before any run. Every send was one model-free decision at the host's egress proxy, and the framework inside the container was unmodified.

\end{document}